\documentclass[11pt]{article}
\newcommand{\be}{\begin{equation}}
\newcommand{\ee}{\end{equation}}
\newcommand{\bea}{\begin{eqnarray}}
\newcommand{\eea}{\end{eqnarray}}
\newcommand{\sn}{{\rm sn}}
\newcommand{\cn}{{\rm cn}}
\newcommand{\dn}{{\rm dn}}

\newcommand{\sech}{{\rm sech}}

\begin{document}

\vspace{0.5in}
\begin{center}
{\LARGE{\bf 
Nonlocal Superposed Solutions II:  
Coupled Nonlinear Equations}}
\end{center}

\begin{center}
{\LARGE{\bf Avinash Khare}} \\
{Physics Department, Savitribai Phule Pune University \\
 Pune 411007, India}
\end{center}

\begin{center}
{\LARGE{\bf Avadh Saxena}} \\ 
{Theoretical Division and Center for Nonlinear Studies, 
Los Alamos National Laboratory, Los Alamos, New Mexico 87545, USA}
\end{center}

\vspace{0.9in}
\noindent{\bf {Abstract:}}

We obtain novel periodic as well as hyperbolic solutions of an 
Ablowitz-Musslimani variant of the coupled nonlocal, nonlinear 
Schr\"odinger equation (NLS) as well as a coupled nonlocal modified 
Korteweg-de Vries (mKdV) equation which can be re-expressed as a 
linear superposition of the sum or the difference of two hyperbolic or 
two periodic kink or pulse solutions. Besides, we also discuss some 
of the other solutions admitted by these coupled equations.
These results demonstrate that the notion of the superposed solutions 
extends to the coupled nonlocal nonlinear equations as well. 

\section{Introduction}
 
Nonlinear theories are highly nontrivial because unlike the 
linear theories, the nonlinear theories do not satisfy the 
superposition principle. However, by now several
nonlinear equations such as $\phi^4$, NLS, mKdV, KdV  etc. have been shown
to satisfy a kind of superposition principle \cite{ks13,ks14}. In particular,
we showed that if a nonlinear equation admits a periodic pulse solution in terms
of the Jacobi elliptic functions \cite{as} $\dn(x,m)$ and $\cn(x,m)$, then the
same equation will also admit a superposed periodic solution 
$\dn(x,m) \pm \sqrt{m} \cn(x,m)$
where $m$ is the modulus of the Jacobi elliptic function. Similarly, if a
nonlinear equation admits a solution in terms of $\dn^2(x,m)$ then it will
also admit a solution in terms of 
$\dn^2(x,m) \pm \sqrt{m} \dn(x,m) \cn(x,m)$. Further, in 
another publication \cite{ks16} we showed that if a nonlinear equation
satisfies solutions in terms of $\dn(x,m)$ or $\cn(x,m)$, then the same
nonlinear equation will also admit a complex PT-invariant solution in
terms of $\dn(x,m) \pm i \sqrt{m} \sn(x,m)$ or 
$\cn(x,m) \pm i \sn(x,m)$. Similarly, if a nonlinear equation admits a 
periodic kink solution in terms of $\sn(x,m)$, then the same nonlinear
equation also admits a complex PT-invariant solution in terms of 
$\sqrt{m} \sn(x,m) \pm i \dn(x,m)$ as well as 
$\sn(x,m) \pm i \cn(x,m)$. We further showed that if a nonlinear
equation admits a solution in terms of $\dn^2(x,m)$ then the same
nonlinear equation will also admit complex PT-invariant periodic
solutions in terms of $\dn^2(x,m) \pm i \sqrt{m} \dn(x,m) \sn(x,m)$
or $\dn^2(x,m) \pm im\sn(x,m) \cn(x,m)$. 

Recently, inspired
by the work of \cite{tan, sma} as well as earlier work in condensed matter
as well as field theory \cite{dashen, campbell, saxena, thies}, 
we  have further extended the notion of superposition for several 
nonlinear equations. 
In particular, we showed that the symmetric and the asymmetric $\phi^4$ model,
the NLS, quadratic-cubic NLS as well as mKdV and mKdV-KdV equations 
admit novel solutions which can be re-expressed as the superposition 
of the periodic kink and the periodic pulse solutions \cite{ks22a}. Further,
in some cases we also obtained solutions which can be re-expressed 
as superposition of two hyperbolic kink 
solutions. In another publication, we further extended the notion of 
the superposition principle for the coupled
nonlinear equations. In particular, we showed that the coupled $\phi^4$,
coupled NLS and coupled mKdV equations admit novel solutions which
can be re-expressed as the superposition of two (hyperbolic) kink and
two (hyperbolic) pulse solutions \cite{ks22b}. 

At this stage we recall that during the last few years the nonlocal, nonlinear
equations have received wide attention. The next obvious question 
is whether this notion can also be extended to nonlocal nonlinear equations. As a 
first step in that direction we recently showed \cite{ks22c} that two 
different nonlocal versions of the NLS \cite{abm1,yang}, nonlocal 
mKdV \cite{he18} as well as the nonlocal Hirota equation \cite{ccf, xyx} 
admit novel solutions which can be re-expressed as the
superposition of two kink or two pulse solutions. It is then natural to
enquire if the notion of superposition can also be extended to the coupled
nonlocal equations. The purpose of this paper is to partially answer
this question. In particular, we show that the Ablowitz-Musslimani
variant \cite{abm1, abm2} of the coupled nonlocal NLS equations \cite{ks14}
as well as the coupled nonlocal 
mKdV equations admit novel solutions which can be re-expressed as the
superposition of two hyperbolic as well as two periodic kink and  
pulse solutions.

The plan of the paper is the following. In Sec. II, we consider a coupled 
Ablowitz-Musslimani variant of the nonlocal NLS and show that it admits a 
large number of novel solutions which can be 
re-expressed as the superposition of two kink or 
two pulse solutions. In particular, while in Sec. IIA we discuss the 
superposed hyperbolic kink or pulse solutions, in Sec. IIB we discuss 
superposed periodic kink or the pulse solutions. 
In Sec. III we discuss a coupled nonlocal mKdV model 
and show that this model also admits novel solutions which can be 
re-expressed as superposition of either two kink or two 
pulse solutions. Finally, in Sec. IV we summarize  
our main results and point out some of the open problems. 

In Appendix A
we discuss those solutions of the coupled Ablowitz-Musslimani 
variant of the nonlocal NLS equation where the two fields are 
proportional to each other. In Appendix B
we discuss those solutions of the coupled nonlocal mKdV model
where the two fields are proportional to each other.

\section{A Coupled Ablowitz-Musslimani Variant of Nonlocal NLS Model}

Let us consider the following coupled Ablowitz-Musslimani variant of the
nonlocal NLS which we had introduced in \cite{ks14}
\be\label{1}
iu_{x,t} +u_{xx}(x,t) +[g_{11} u(x,t)u^{*}(-x,t)+g_{12} v(x,t) v^{*}(-x,t)]
u(x,t) = 0\,,
\ee
\be\label{2} 
iv_{x,t} +v_{xx}(x,t) +[g_{21} u(x,t)u^{*}(-x,t)+g_{22} v(x,t) v^{*}(-x,t)]
v(x,t) = 0\,. 
\ee
Note that for comparison with the standard version of the Manakov system
\cite{man}, 
we have changed the notation used in \cite{ks14}. In particular, we have
replaced $z$ with $t$ and also instead of the couplings $a,b,f,e$ as used in
\cite{ks14} we have used the couplings $g_{11}, g_{12}, g_{21}$ and $g_{22}$, 
respectively. Notice that the limit $g_{11}=g_{12}=g_{21}=g_{22}$ in the local
coupled NLS case is the Manakov system while $g_{11}=g_{21}=-g_{12}=-g_{22}$
in the local case is the integrable MZS system \cite{mik,zs,ger}. 

In \cite{ks14} we had obtained 34 solutions of the coupled 
equations, most of which were periodic in terms of the Jacobi elliptic 
functions and a few in terms of hyperbolic solutions. However, it turns out 
that we had missed several interesting solutions of the coupled equations. 
The purpose of this paper is to discuss these solutions and 
to show that they can be re-expressed as the superposition of a kink and
an antikink or two kink or two pulse solutions.
For this purpose we will make use of six 
identities for the Jacobi elliptic functions which can be easily derived by 
using the well known addition theorems for the Jacobi elliptic 
functions $\sn(x,m), \cn(x,m)$ and $\dn(x,m)$ \cite{as} 
\be\label{3}
\sn(a+b,m) = \frac{\sn(a,m)\cn(b,m)\dn(b,m)+\sn(b,m)\cn(a,m)\dn(a,m)}
{1-m\sn^2(a,m)\sn^2(b,m)}\,,
\ee
\be\label{4}
\cn(a+b,m) = \frac{\cn(a,m)\cn(b,m)-\sn(a,m)\dn(a,m)\sn(b,m)\dn(b,m)}
{1-m\sn^2(a,m)\sn^2(b,m)}\,,
\ee
\be\label{5}
\dn(a+b,m) = \frac{\dn(a,m)\dn(b,m)-m \sn(a,m)\cn(a,m)\sn(b,m)\cn(b,m)}
{1-m\sn^2(a,m)\sn^2(b,m)}\,.
\ee
In particular, on using Eq. (\ref{3}) we obtain the identities for the
sum of two periodic kink solutions, i.e.
\be\label{6}
\sn(y+\Delta,m)+\sn(y-\Delta,m) = \frac{2\sn(y,m) \cn(\Delta,m)}
{\dn(\Delta,m)[1+B\cn^2(y,m)]}\,,
~~B = \frac{m\sn^2(\Delta,m)}{\dn^2(\Delta,m)}\,, 
\ee
or the sum of a periodic kink and an antikink solution, i.e.
\be\label{7}
\sn(y+\Delta,m)-\sn(y-\Delta,m) = \frac{2\cn(y,m) \dn(y,m)\sn(\Delta,m)}
{\dn^2(\Delta,m)[1+B\cn^2(y,m)]}\,.
~~B = \frac{m\sn^2(\Delta,m)}{\dn^2(\Delta,m)}\,. 
\ee
On the other hand, by using Eq. (\ref{4}) we obtain the identities
for the sum of the two periodic pulse solutions, i.e. 
\be\label{8}
\cn(y+\Delta,m)+\cn(y-\Delta,m) = \frac{2\cn(y,m) \cn(\Delta,m)}
{\dn^2(\Delta,m)[1+B\cn^2(y,m)]}\,,
~~B = \frac{m\sn^2(\Delta,m)}{\dn^2(\Delta,m)}\,, 
\ee
\be\label{9}
\cn(y-\Delta,m)-\cn(y+\Delta,m) = \frac{2\sn(y,m) \dn(y,m)\sn(\Delta,m)}
{\dn(\Delta,m)[1+B\cn^2(y,m)]}\,.
~~B = \frac{m\sn^2(\Delta,m)}{\dn^2(\Delta,m)}\,. 
\ee
Finally, on using the addition theorem (\ref{5}) we obtain the identities
for the sum of the two periodic pulse solutions, i.e. 
\be\label{10}
\dn(y+\Delta,m)+\dn(y-\Delta,m) = \frac{2\dn(y,m)}
{\dn(\Delta,m)[1+B\cn^2(y,m)]}\,,
~~B = \frac{m\sn^2(\Delta,m)}{\dn^2(\Delta,m)}\,,  
\ee
\bea\label{11}
&&\dn(y-\Delta,m)-\dn(y+\Delta,m) = \frac{2m \sn(y,m) \cn(y,m)
\sn(\Delta,m) \cn(\Delta,m)}{\dn^2(\Delta,m)[1+B\cn^2(y,m)]}\,,
 \nonumber \\
&&B = \frac{m\sn^2(\Delta,m)}{\dn^2(\Delta,m)}\,. 
\eea

In the limit $m = 1$, from the above identities we can deduce the 
corresponding four hyperbolic identities. In particular, in the limit $m = 1$, 
from Eq. (\ref{6}) we obtain an identity for the sum of two
(hyperbolic) kink solutions, i.e.
\be\label{13}
\tanh(y+\Delta)+\tanh(y-\Delta) = \frac{2\sinh(y)\cosh(y)}
{[B+\cosh^2(y)]}\,,~~B = \sinh^2(\Delta)\,,
\ee
while from Eq. (\ref{7}) in the limit $m = 1$ we obtain an identity
for the sum of a (hyperbolic) kink and an antikink solution, i.e.
\be\label{14}
\tanh(y+\Delta)-\tanh(y-\Delta) = \frac{\sinh(2\Delta)}
{[B+\cosh^2(y)]}\,,~~B = \sinh^2(\Delta)\,.
\ee
On the other hand, from Eqs. (\ref{8}) and (\ref{9}) (or (\ref{10})
and (\ref{11})), in the $m = 1$ limit, we obtain the identities
for the sum of two (hyperbolic) pulse solutions
\be\label{15}
\sech(y+\Delta)+\sech(y-\Delta) = \frac{2\cosh(y) \cosh(\Delta)}
{[B+\cosh^2(y)]}\,,~~B = \sinh^2(\Delta)\,, 
\ee
\be\label{16}
\sech(y-\Delta)-\sech(y+\Delta) = \frac{2 \sinh(y)\sinh(\Delta)}
{[B+\cosh^2(y)]}\,,~~B = \sinh^2(\Delta)\,.
\ee

In the next two subsections, we obtain 7 hyperbolic and 19 periodic 
solutions, respectively, of the coupled 
Eqs. (\ref{1}) and (\ref{2}) which can be re-expressed as the
superposition of either a kink and an antikink or two kink or 
two pulse solutions. 

\subsection{Superposed Hyperbolic kink and Pulse Solutions}

There are two different kind of solutions to the coupled Eqs. (\ref{1})
and (\ref{2}) depending on whether $v(x,t) \propto u(x,t)$ or not. 
In this section we only discuss those solutions where the two fields 
$v(x,t)$ and $u(x,t)$ are distinct (and not proportional to each other) 
while in Appendix A we discuss those solutions where $v(x,t) \propto
u(x,t)$. As we show there, when the two fields $u$ and $v$ are 
proportional to each other, effectively we only need to solve the 
uncoupled Ablowitz-Musslimani nonlocal NLS equation.

We start from Eqs. (\ref{1}) and (\ref{2}) and take the ansatz
\be\label{1.1}
u(x,t) = e^{i\omega_1 (t+t_0)} u(x)\,,~~v(x,t) 
= e^{i\omega_2 (t+t_0)} v(x)\,.
\ee
On substituting this ansatz in Eqs. (\ref{1}) and (\ref{2}) we obtain
\be\label{1.2}
u_{xx}(x) = \omega_1 u(x) -[g_{11} u(x)u^{*}(-x)+g_{12} v(x) v^{*}(-x)]
u(x)\,,
\ee
\be\label{1.3} 
v_{xx}(x) = \omega_2 v(x) -[g_{21} u(x)u^{*}(-x)+g_{22} v(x) v^{*}(-x)]
v(x)\,.
\ee

We now show that the coupled Eqs. (\ref{1.2}) and (\ref{1.3}) and hence
Eqs. (\ref{1}) and (\ref{2}) admit 7 novel hyperbolic solutions which
can be re-expressed as a superposition of either two kink or two pulse
solutions.

One major difference between the local and the nonlocal case is that, 
unlike the local case, the solutions of the nonlocal NLS Eqs. (\ref{1}) and 
(\ref{2}) are not invariant with respect to the shift in $x$. However, a shift 
 in $t$ is allowed. For simplicity, from now onward we will take $t_0$ to  
 be zero even though such a shift is always allowed.

{\bf Superposed Solution I}

It is easy to check that the coupled Eqs. (\ref{1.2}) and 
(\ref{1.3}) (and hence Eqs. (\ref{1}) and (\ref{2})) admit the 
hyperbolic solution
\be\label{1.4}
u(x,t) = e^{i\omega_1 t} \frac{A}{B+\cosh^2(\beta x)}\,,
~~v(x,t) = e^{i\omega_2 t} \frac{D \cosh(\beta x) \sinh(\beta x)}
{B+\cosh^2(\beta x)}\,,~~B > 0\,,
\ee
provided
\bea\label{1.5}
&&\omega_1 = -2\beta^2\,,~~ g_{11} A^2 
= -2B(2B+1)\beta^2\,,~~g_{12} D^2 = 6\beta^2\,, \nonumber \\
&&\omega_2 = -2\beta^2\,,~~g_{21} A^2 = -6B(B+1) \beta^2\,,
~~g_{22} D^2 = 2 \beta^2\,.
\eea
Thus for this solution $g_{11}, g_{21}$ are negative while 
$\omega_1 = \omega_2$, $g_{12}$, $g_{22}$ are positive.

On comparing the solution (\ref{1.4}) with the identities (\ref{13}) 
and (\ref{14}), the solution (\ref{1.4}) can be
re-expressed as
\bea\label{1.6}
&&u(x,t) = e^{i\omega_1 t} \frac{\beta}{\sqrt{2|g_{11}|}}
[\tanh(\beta x + \Delta) -\tanh(\beta x - \Delta)]\,, 
\nonumber \\
&&v(x,t) = e^{i\omega_2 t} \frac{\sqrt{3}\beta}{\sqrt{2g_{12}}}
[\tanh(\beta x + \Delta)+\tanh(\beta x - \Delta)]\,,
\eea
where $\sinh^2(\Delta) = B$. Note that here while $u(x,t)$ corresponds to a 
superposition of a kink and an antikink solution, $v(x,t)$ is a
superposition of two kink solutions. 

{\bf Superposed Solution II}

It is easy to check that the coupled Eqs. (\ref{1}) and 
(\ref{2}) admit the superposed hyperbolic solution
\be\label{1.7}
u(x,t) = e^{i\omega_1 t} \frac{A}{B+\cosh^2(\beta x)}\,,
~~v(x,t) = e^{i\omega_2 t} \frac{D \sinh(\beta x)}
{B+\cosh^2(\beta x)}\,,~~B > 0\,,
\ee
provided
\bea\label{1.8}
&&\omega_1 = 4\beta^2\,,~~g_{11} A^2 = 2(B+1)(2B+3)\beta^2\,,
~~g_{12} D^2 = -6(2B+1)\beta^2\,, \nonumber \\
&&\omega_2 = \beta^2\,,~~g_{21} A^2 = 6(B+1) \beta^2\,,
~~g_{22} D^2 = -2(3+4B)\beta^2\,.
\eea
Thus for this solution $g_{11}$, $g_{21} $, $\omega_1$, $\omega_2$ 
are all positive while $g_{12}$, $g_{22}$ are negative.

On comparing the solution (\ref{1.7}) with the identities (\ref{14}) 
and (\ref{16}), the solution (\ref{1.7}) can be re-expressed as
\bea\label{1.9}
&&u(x,t) = e^{i\omega_1 t} \frac{\sqrt{3}\beta}
{\sqrt{2g_{21}} \sinh(\Delta)}
[\tanh(\beta x + \Delta) -\tanh(\beta x - \Delta)]\,, \nonumber \\
&&v(x,t) = e^{i\omega_2 t} \frac{\sqrt{3 \cosh(2\Delta)}\beta}
{\sqrt{2 |g_{12}|}\sinh(\Delta)}
[\sech(\beta x - \Delta)- \sech(\beta x +\Delta)]\,,
\eea
where $\sinh^2(\Delta) = B$. Note that here while $u(x,t)$ corresponds to a 
superposition of a kink and an antikink solution, $v(x,t)$ is a
superposition of two pulse solutions. 

{\bf Superposed Solution III}

It is easy to check that the coupled Eqs. (\ref{1}) and 
(\ref{2}) admit the superposed hyperbolic solution
\be\label{1.10}
u(x,t) = e^{i\omega_1 t} \frac{A}{B+\cosh^2(\beta x)}\,,
~~v(x,t) = e^{i\omega_2 t} \frac{D \cosh(\beta x)}
{B+\cosh^2(\beta x)}\,,~~B > 0\,,
\ee
provided
\bea\label{1.11}
&&\omega_1 = 4\beta^2\,,~~g_{12} D^2 = 6(1+2B)\beta^2\,,
~~g_{11} A^2 = 2B(2B-1)\beta^2\,, \nonumber \\
&&\omega_2 = \beta^2\,,~~g_{21} A^2 = -6 B \beta^2\,,
~~g_{22} D^2 = 2(1+4B)\beta^2\,.
\eea
Thus for this solution while $g_{12}$, $g_{22}$, $\omega_1$, $\omega_2  > 0$, 
$g_{21} < 0$, $g_{11} \ge 0$ if either $B \ge  1/2$ or $B \le 0$,  
and $g_{11} < 0$ if $0 < B < 1/2$. 

On comparing the solution (\ref{1.10}) with the identities (\ref{14}) 
and (\ref{15}), the solution (\ref{1.10}) can be re-expressed as
\bea\label{1.12}
&&u(x,t) = e^{i\omega_1 t} 
\frac{\sqrt{3}\beta} {\sqrt{2|g_{21}|} \cosh(\Delta)}
[\tanh(\beta x + \Delta) -\tanh(\beta x - \Delta)]\,, 
\nonumber \\
&&v(x,t) =  e^{i\omega_2 t} 
\frac{\sqrt{3} \tanh(\Delta)\beta}{\sqrt{2 |g_{21}|}}
[\sech(\beta x + \Delta) + \sech(\beta x - \Delta)]\,,
\eea
where $\sinh^2(\Delta) = B$. Note that here while $u(x,t)$ corresponds to a 
superposition of a kink and an antikink solution, $v(x,t)$ is a
superposition of two pulse solutions. 

{\bf Superposed Solution IV}

It is easy to check that the
coupled Eqs. (\ref{1}) and (\ref{2}) admit the superposed 
hyperbolic solution
\be\label{1.13}
u(x,t) = e^{i\omega_1 t} \frac{A\sinh(\beta x)}
{B+\cosh^2(\beta x)}\,,
~~v(x,t) = e^{i\omega_2 t} \frac{D \sinh(\beta x) \cosh(\beta x)}
{B+\cosh^2(\beta x)}\,,~~B > 0\,,
\ee
provided
\bea\label{1.14}
&&(B+1)\omega_1 = -(5-B)\beta^2\,,~~(1+B) g_{11} A^2 
= -2 B(1+4B) \beta^2\,, \nonumber \\
&&(1+B) g_{12} D^2 = 6 \beta^2\,, ~~(1+B) \omega_2 = 2(2B-1) \beta^2\,,
 \nonumber \\
&&(B+1) g_{21} A^2 = -6B(2B+1) \beta^2\,,~~
(B+1) g_{22} D^2 = -2(2B-1)\beta^2\,. ~~~
\eea
Thus for this solution $g_{11}$, $g_{21} < 0$ while $g_{12} > 0$. 

On making use of the identities (\ref{13}) and (\ref{16}), one
can then re-express the solution (\ref{1.13}) as
\bea\label{1.15}
&&u(x,t) = e^{i\omega_1 t} \frac{\beta}
{\sqrt{2 |g_{11}|}}
[\sech(\beta x - \Delta) -\sech(\beta x + \Delta)]\,, \nonumber \\
&&v(x,t) = e^{i\omega_2 t} \frac{\sqrt{3}\beta}
{\sqrt{2 g_{12}}\cosh(\Delta)} 
[\tanh(\beta x + \Delta) + \tanh(\beta x -\Delta)]\,,
\eea
where $\sinh^2(\Delta) = B$. Note that here while $u(x,t)$ corresponds to a 
superposition of two pulse solutions, $v(x,t)$ is a superposition of two 
kink solutions.
\vskip 0.4cm 

{\bf Superposed Solution V}

It is easy to check that the coupled Eqs. (\ref{1}) and 
(\ref{2}) admit the superposed hyperbolic solution
\be\label{1.16}
u(x,t) = e^{i\omega_1 t} \frac{A\cosh(\beta x)}
{B+\cosh^2(\beta x)}\,,~~v(x,t) = e^{i\omega_2 t} 
\frac{D \sinh(\beta x) \cosh(\beta x)}{B+\cosh^2(\beta x)}\,,~~B > 0\,,
\ee
provided
\bea\label{1.17}
&&B \omega_1 = (6+B)\beta^2\,,~~B g_{11} A^2 = 
2(B+1)(4B+3)\beta^2\,, \nonumber \\
&&B g_{12} D^2 = -6 \beta^2\,,~~
B \omega_2 = 2(2B+3)\beta^2\,, \nonumber \\
&&B g_{21} A^2 = 6(B+1)(2B+1)\beta^2\,,
~~B g_{22} D^2 = -2(3+2B)\beta^2\,.
\eea
Thus for this solution $g_{11}$, $g_{21}$, $\omega_1$, $\omega_2$ 
are all positive while $g_{12}$, $g_{22}$ are negative.

On making use of the identities (\ref{13}) and (\ref{15}), one
can then re-express the solution (\ref{2.26}) as
\bea\label{1.18}
&&u(x,t) = e^{i\omega_1 t} \frac{\sqrt{3\cosh(2\Delta)}\beta}
{\sqrt{2 g_{21}}\sinh(\Delta)}
[\sech(\beta x + \Delta) +\sech(\beta x - \Delta)]\,, 
\nonumber \\
&&v(x,t) = e^{i \omega_2 t} \frac{\sqrt{3}\beta}
{\sqrt{2 |g_{12}|}\sinh(\Delta)}
[\tanh(\beta x + \Delta)+ \tanh(\beta x -\Delta)]\,,
\eea
where $\sinh^2(\Delta) = B$. Note that here while $u(x,t)$ corresponds to a 
superposition of two pulse solutions, $v(x,t)$ is a superposition of two 
kink solutions.

{\bf Superposed Solution VI}

It is easy to check that the coupled Eqs. (\ref{1}) and 
(\ref{2}) admit the superposed hyperbolic solution
\be\label{1.19}
u(x,t) = e^{i\omega_1 t} \frac{A\sinh(\beta x)}
{B+\cosh^2(\beta x)}\,,~~v(x,t) = e^{i\omega_2 t} 
\frac{D \cosh(\beta x)}{B+\cosh^2(\beta x)}\,,~~B > 0\,,
\ee
provided
\bea\label{1.20}
&&\omega_1 = \beta^2\,,~~g_{11} A^2 = -2B \beta^2\,,
~~g_{12} D^2 = 6(B+1)\beta^2\,, \nonumber \\
&&\omega_2 = \beta^2\,,~~g_{21} A^2 = -6 B \beta^2\,,
~~g_{22} D^2 = 2(B+1)\beta^2\,.
\eea
Thus for this solution $g_{12}$, $g_{22}$, $\omega_1$, $\omega_2$ 
are all positive while $g_{11}$, $g_{21}$ are negative.

On making use of the identities (\ref{15}) and (\ref{16}), one
can then re-express the solution (\ref{1.19}) as
\bea\label{1.21}
&&u(x,t) = e^{i \omega_1 t} \frac{\beta}{\sqrt{2 |g_{11}|}}
[\sech(\beta x - \Delta) -\sech(\beta x + \Delta)]\,, \nonumber \\
&&v(x,t) = e^{i \omega_2 t} \frac{\beta}{\sqrt{2 g_{22}}}
[\sech(\beta x + \Delta)+ \sech(\beta x -\Delta)]\,,
\eea
where $\sinh^2(\Delta) = B$. Note that here both $u(x,t), v(x,t)$  
correspond to the superposition of two pulse solutions.

{\bf Superposed Solution VII}

It is easy to check that the coupled Eqs. (\ref{1}) and 
(\ref{2}) admit the superposed hyperbolic solution
\be\label{1.22}
u(x,t) = e^{i\omega_1 t} [1- \frac{A}{B+\cosh^2(\beta x)}]\,,
~~v(x,t) = e^{i\omega_2 t} \frac{D}
{B+\cosh^2(\beta x)}\,,
\ee
where $A, B, D > 0$ provided
\bea\label{1.23}
&&g_{11} = \omega_1 = -2\beta^2\,,~~A = \frac{3(1+2B)+\sqrt{8+(2B+1)^2}}{4}\,,
\nonumber \\
~&&g_{12} D^2 = -[\sqrt{8+(2B+1)^2}-(2B+1)] \frac{3A\beta^2}{2}\,,
~~\omega_2 -g_{21} = 4\beta^2\,, \nonumber \\
&&g_{21} A = -3(2B+1) \beta^2\,,
~~g_{21} A^2 + g_{22} D^2 = -8B(B+1) \beta^2\,.
\eea
Thus for this solution $g_{11}$, $g_{12}$, $\omega_1$, $g_{21} <0$, 
while $\omega_2$, $g_{22} >0$.

On comparing the solution (\ref{1.22}) with the identity (\ref{16}), 
the solution (\ref{1.22}) can be re-expressed as
\bea\label{1.24}
&&u(x,t) = e^{i\omega_1 t} \left[1- \frac{K_1 \beta}
{\sinh(2\Delta)\sqrt{2|g_{11}|}}\right]
[\tanh(\beta x + \Delta)-\tanh(\beta x - \Delta)]\,, \nonumber \\
&&v(x,t) = e^{i\omega_2 t} \frac{\sqrt{3K_1 K_3}\beta}
{2\sinh(2\Delta) \sqrt{2 |g_{12}|}} 
[\tanh(\beta x + \Delta)-\tanh(\beta x - \Delta)]\,,
\eea
where $\sinh^2(\Delta) = B$ and 
\be\label{1.25}
K_1 = \left[3\cosh(2\Delta)+\sqrt{8+\cosh^2(2\Delta)}\right]\,, 
\ee
\be\label{1.26}
K_3 = \left[\sqrt{8+\cosh^2(2\Delta)-\cosh(2\Delta)}\right]\,.
\ee
Note that here both $u(x,t), v(x,t)$  
correspond to the superposition of a kink and an antikink solution.

\subsection{Superposed Periodic Kink and Pulse Solutions}

We now show that in case $u$
and $v$ are distinct (i.e. not proportional to each other)
then by starting from the ansatz (\ref{1.1}) and hence using the 
coupled Eqs. (\ref{1.2}) and (\ref{1.3}), and thus the coupled Eqs. 
(\ref{1}) and (\ref{2}) admit not only the 7 superposed
hyperbolic but also 15 superposed periodic kink (i.e. $\sn(x,m)$) and periodic 
pulse (i.e. $\dn(x,m)$ as well as $\cn(x,m)$) solutions. We now present these
15 solutions one by one. 

{\bf Solution VIII}

It is easy to check that
\be\label{2.22}
u(x,t) = e^{i\omega_1 t} \frac{A\cn(\beta x,m)}{1+B\cn^2(\beta x,m)}\,, ~~~
v(x,t) = e^{i\omega_2 t} \frac{D \sn(\beta x,m)\dn(\beta x,m)}
{1+B\cn^2(\beta x,m)}\,,
\ee
where $B > 0$, is an exact solution of the coupled 
Eqs. (\ref{1}) and (\ref{2}) provided
\bea\label{2.23}
&&\omega_1 = \omega_2 = (2m-1)\beta^2\,,~~g_{12} D^2 = 3 g_{22} D^2 
= -6 B\beta^2\,, \nonumber \\
&&g_{21} A^2  = 3 g_{11} A^2 = 6(B+1)[m-(1-m)B]\beta^2\,.
\eea

On using the identities (\ref{8}) and (\ref{9}), the coupled 
solution (\ref{2.22}) can be re-expressed as 
\bea\label{2.24}
&&u(x,t) = e^{i\omega_1 t} \sqrt{\frac{m}{2g_{21}}}\beta  
[\cn(\beta x+\Delta,m)+\cn(\beta x-\Delta,m)]\,, \nonumber \\
&&v(x,t) = e^{i\omega_2 t} \sqrt{\frac{m}{2 |g_{12}|}} \beta
[\cn(\beta x-\Delta,m)-\cn(\beta x+\Delta,m)]\,,
\eea
where $B = {m\sn^2(\Delta,m)}/{\dn^2(\Delta,m)}$. 

{\bf Solution IX}

It is easy to check that
\be\label{2.54}
u(x,t) = e^{i\omega_1 t} \frac{A\dn(\beta x,m)}{1+B\cn^2(\beta x,m)}\,,
~~v(x,t) = e^{i\omega_2 t} \frac{D \sn(\beta x,m) \cn(\beta x,m)}
{1+B\cn^2(\beta x,m)}\,,
\ee
where $B > 0$, is an exact solution of the coupled 
Eqs. (\ref{1}) and (\ref{2}) provided
\bea\label{2.55}
&&\omega_1 = \omega_2 = (2-m)\beta^2\,,~~g_{21} D^2 = 3 g_{22} D^2 
= -6B[m-(1-m)B]\beta^2\,, \nonumber \\
&&g_{21} A^2 = 3 g_{11} A^2 = 6(1+B) \beta^2\,.
\eea

On using the identities (\ref{10}) and (\ref{11}), the coupled 
solution (\ref{2.54}) can be re-expressed as 
\bea\label{2.56}
&&u(x,t) = e^{i\omega_1 t} \frac{\beta}{\sqrt{2 g_{11}}}
[\dn(\beta x+\Delta,m)+\dn(\beta x-\Delta,m)]\,, \nonumber \\
&&v(x,t) = e^{i\omega_2 t} \frac{\beta}
{\sqrt{2 |g_{22}|}}[\dn(\beta x-\Delta,m)-\dn(\beta x+\Delta,m)]\,,
\eea
where $B = {m\sn^2(\Delta,m)}/{\dn^2(\Delta,m)}$. 

{\bf Solution X}

It is straightforward to check that
\be\label{2.16}
u(x,t) = e^{i\omega_1 t} \frac{A\cn(\beta x,m)}{1+B\cn^2(\beta x,m)}\,,
~~v(x,t) = e^{i\omega_2 t} \frac{D\cn(\beta x,m) \sn(\beta x,m)}
{1+B\cn^2(\beta x,m)}\,,
\ee
where $B > 0$, is an exact solution to the coupled 
Eqs. (\ref{1}) and (\ref{2}) provided
\bea\label{2.17}
&&\omega_1 = [(2m-1) -6(1-m)B]\beta^2\,,~~g_{12} D^2 = -6[m+(1-m)B^2]B\beta^2\,,
\nonumber \\
&&g_{11} A^2 = -2[3(1-m)B^3+7(1-m)B^2+B(4-5m)-m]\beta^2 \,, \nonumber \\
&&\omega_2 = [(5m-4)-6(1-m)B]\beta^2\,, \nonumber \\
&&g_{22} D^2 = -2[3(1-m)B^2+2(1-m)B+m]B\beta^2\,, \nonumber \\
&&b_{21} A^2 = -6[(1-m)B^3+3(1-m)B^2 +(2-3m)B-m]\beta^2\,.
\eea

On using the identities (\ref{8}) and (\ref{11}), the coupled 
solution (\ref{2.16}) can be re-expressed as 
\be\label{2.18}
u(x,t) = e^{i\omega_1 t} \frac{A \dn^2(\Delta,m)}{2\cn(\Delta,m)} 
[\cn(\beta x+\Delta,m)+\cn(\beta x-\Delta,m)]\,,
\ee
\be\label{2.18a}
v(x,t) = e^{i\omega_2 t} \frac{D \dn^2(\Delta)}{2m\sn(\Delta)\cn(\Delta)}
[\dn(\beta x-\Delta,m)-\dn(\beta x+\Delta,m)]\,,
\ee
where $A, D$ are as given by Eq. (\ref{2.17}) while 
$B = {m\sn^2(\Delta,m)}/{\dn^2(\Delta,m)}$. 

{\bf Solution XI}

It is easy to check that
\be\label{2.29}
u(x,t) = e^{i\omega_1 t} \frac{A\dn(\beta x,m)}{1+B\cn^2(\beta x,m)}\,,
~~v(x,t) = e^{i\omega_2 t} \frac{D \sn(\beta x,m)\dn(\beta x,m)}
{1+B\cn^2(\beta x,m)}\,,
\ee
where $B > 0$, is an exact solution of the coupled 
Eqs. (\ref{1}) and (\ref{2}) provided
\bea\label{2.30}
&&[m-(1-m)B] \omega_1 = [(1-m)(4+m)B+m(2-m)]\beta^2\,, \nonumber \\
&&[m-(1-m)B] g_{11} A^2 = 2(B+1)[3(1-m)B^2+2(1-m)B+m]\,, 
\nonumber \\
&&m[m-(1-m)B] g_{12} D^2 = -6B[m- 2(1-m)B-(1-m)B^2]\beta^2\,, \nonumber \\
&&[m-(1-m)B] \omega_2 = [(1-m)(4m+1)B+m(5-4m)]\beta^2\,, \nonumber \\
&&m [m-(1-m)B] g_{22} D^2 = -B[2m^2-5m(1-m)B -3(1-m^2) B^2]\beta^2\,,
\nonumber \\
&&[m-(1-m)B] g_{21} A^2  = [6(1-m)B^3 +6(1-m)B^2+(m^2+4m+1)B+6m] \beta^2\,. 
\eea

On using the identities (\ref{9}) and (\ref{10}), the coupled 
solution (\ref{2.29}) can be re-expressed as 
\be\label{2.31}
u(x,t) = e^{i\omega_1 t} \frac{A \dn(\Delta,m)}{2}
[\dn(\beta x+\Delta,m)+\dn(\beta x-\Delta,m)]\,,
\ee
\be\label{2.31a}
v(x,t) = e^{i\omega_2 t} \frac{D \dn(\Delta,m)}{2\sn(\Delta,m)}
[\cn(\beta x-\Delta,m)-\cn(\beta x+\Delta,m)]\,, 
\ee
where $A, D$ are given by Eq. (\ref{2.30}) while 
$B = {m\sn^2(\Delta,m)}/{\dn^2(\Delta,m)}$.

{\bf Hyperbolic Limit}

In the limit $m = 1$, all the four solutions VIII to XI go over to the
hyperbolic solution VI, i.e. in this limit all four solutions can 
be re-expressed as superposition of two pulse solutions, i.e. 
$\sech(\beta x- \Delta) \pm \sech(\beta x +\Delta)$.

{\bf Solution XII}

It is straightforward to check that
\be\label{2.32}
u(x,t) = e^{i\omega_1 t}\frac{A\cn(\beta x,m)}{1+B\cn^2(\beta x,m)}\,,
~~v(x,t) = e^{i\omega_2 t} \frac{D \sn(\beta x,m)}{1+B\cn^2(\beta x,m)}\,,
\ee
where $B > 0$, is an exact solution of the coupled 
Eqs. (\ref{1}) and (\ref{2}) provided
\bea\label{2.33a}
&&B \omega_1 = [(2m-1)B+6m]\beta^2\,,~~
B g_{12} D^2 =  -6[m+(1-m)B]\beta^2\,, \nonumber \\
&&B g_{11} A^2 =  2[3m+7mB+(5m-1)B^2-(1-m)B^3] 
B \omega_2 = [(5m-1)B+6m]\beta^2\,, \nonumber \\
&&B g_{22} D^2 = -2[3m+2m B +(1-m) B^2]\beta^2\,,~~
B g_{21} A^2  = 6(B+1)[m+2mB-(1-m)B^2]\,. 
\eea

On using the identities (\ref{6}) and (\ref{8}), the coupled 
solution (\ref{2.32}) can be re-expressed as 
\be\label{2.32a}
u(x,t) = e^{i\omega_1 t}\frac{A \dn^2(\Delta,m)}{2\cn(\Delta,m)}
[\cn(\beta x+\Delta,m)+\cn(\beta x-\Delta,m)]\,,
\ee
\be\label{2.32b}
v(x,t) = e^{i\omega_2 t} \frac{D \dn(\Delta,m)}{2\cn(\Delta,m)}
[\sn(\beta x+\Delta,m)+\sn(\beta x-\Delta,m)]\,,
\ee
where $A, D$ are given by Eq. (\ref{2.33a}) while 
$B = {m\sn^2(\Delta,m)}/{\dn^2(\Delta,m)}$. 

{\bf Solution XIII}

It is easy to check that
\be\label{2.36}
u(x,t) = e^{i\omega_1 t} \frac{A\dn(\beta x,m)}{1+B\cn^2(\beta x,m)}\,,
~~~v(x,t) = e^{i\omega_2 t} \frac{D \sn(\beta x,m)}{1+B\cn^2(\beta x,m)}\,,
\ee
where $B > 0$, is an exact solution of the coupled 
Eqs. (\ref{1}) and (\ref{2}) provided
\bea\label{2.37}
&&B \omega_1 = [(5m-4)B+6] \beta^2\,,
~~B \omega_2 = [((5m-1)B+6m] \beta^2\,, \nonumber \\ 
&&B g_{12} D^2 
= -6[(1-m)^2 B^3+(1-m)(2-3m)B^2-3m(1-m)B+m^2] \beta^2\,, 
\nonumber \\
&&B g_{11} A^2 = 2(B+1)[3m +2(3m-1)B-3(1-m)B^2]\beta^2\,, 
\beta^2\,, \nonumber \\
&&B g_{21} A^2  = 6(B+1)[m+2m B-(1-m)B^2]\beta^2\,,
\nonumber \\
&&B g_{22} D^2 = -2[3m^2+m(9m-7)B +(4-9m)(1-m) B^2 +3(1-m)^2] \,. 
\eea

On using the identities (\ref{6}) and (\ref{10}), the coupled 
solution (\ref{2.36}) can be re-expressed as 
\be\label{2.38}
u(x,t) = e^{i\omega_1 t} \frac{A \dn(\Delta,m)}{2}
[\dn(\beta x+\Delta,m)+\dn(\beta x-\Delta,m)]\,, 
\ee
\be\label{2.38a}
v(x,t) = e^{i\omega_2 t} \frac{D \dn(\Delta,m)}{2\cn(\Delta,m)}
[\sn(\beta x+\Delta,m)+\sn(\beta x-\Delta,m)]\,,
\ee
where $A, D$ are given by Eq. (\ref{2.37}) while 
$B = {m\sn^2(\Delta,m)}/{\dn^2(\Delta,m)}$.

{\bf Hyperbolic Limit}

In the limit $m = 1$, the two solutions XII and XIII go over to the
hyperbolic solution V, i.e. in this limit both the solutions can 
be re-expressed as superposition of pulse   
(i.e. $\sech(\beta x+\Delta) + \sech(\beta x-\Delta)$) and 
kink (i.e. $\tanh(\beta x +\Delta) + \tanh(\beta x-\Delta)$) 
solutions, respectively.

{\bf Solution XIV}

It is straightforward to check that
\be\label{2.33}
u(x,t) = e^{i\omega_1 t} \frac{A\cn(\beta x,m)}{1+B\cn^2(\beta x,m)}\,,
~~v(x,t) = e^{i\omega_2 t} \frac{D \cn(\beta x,m)\dn(\beta x,m)}
{1+B\cn^2(\beta x,m)}\,,
\ee
where $B> 0$, is an exact solution of the coupled 
Eqs. (\ref{1}) and (\ref{2}) provided
\bea\label{2.34}
&&\omega_1 = [(2m-1)-6(1-m)B]\beta^2\,,~~
m g_{12} D^2 = -6B[m-(1-m)B^2]\beta^2\,, \nonumber \\
&&m g_{11} A^2 = 2[m^2+m(5m-1)B-7m(1-m)B^2-3(1-m)^2 B^3] \,, \nonumber \\
&&\omega_2 = [((5m-1)-6(1-m)B]\beta^2\,,~~
m g_{22} D^2 = 6[m+2m B +3(1-m) B^2]\beta^2\,, \nonumber \\
&&m g_{21} A^2  = 2B[-m^2+(3m-1) mB-3m(1-m)B^2+(1-m)^2 B^3]\beta^2\,.
\eea

On using the identities (\ref{7}) and (\ref{8}), the coupled 
solution (\ref{2.33}) can be re-expressed as 
\be\label{2.35}
u(x,t) = e^{i\omega_1 t} \frac{A \dn^{2}(\Delta,m)}{2\cn(\Delta,m)}
[\cn(\beta x+\Delta,m)+\cn(\beta x-\Delta,m)]\,, 
\ee
\be\label{2.36a}
v(x,t) = e^{i\omega_2 t} \frac{D \dn^2(\Delta,m)}{2\sn(\Delta,m)}
[\sn(\beta x+\Delta,m)-\sn(\beta x-\Delta,m)]\,,
\ee
where $A, D$ are given by Eq. (\ref{2.34}) while 
$B = {m\sn^2(\Delta,m)}/{\dn^2(\Delta,m)}$. 

{\bf Solution XV}

It is easy to check that
\be\label{2.60}
u(x,t) = e^{i\omega_1 t} \frac{A\dn(\beta x,m)}{1+B\cn^2(\beta x,m)}\,,
~~v(x,t) = e^{i\omega_2 t} \frac{D \cn(\beta x,m) \dn(\beta x,m)}
{1+B\cn^2(\beta x,m)}\,,
\ee
where $B > 0$, is an exact solution of the coupled 
Eqs. (\ref{1}) and (\ref{2}) provided
\bea\label{2.61}
&&[m-(1-m)B] \omega_1 = [m(2-m)+(1-m)(4+m)B]\beta^2\,,
\nonumber \\
&&[m-(1-m)B] g_{11} A^2 = 2[m+2(1+m)B-(1-m)B^2] \beta^2\,, 
\nonumber \\
&&[m-(1-m)B] g_{12} D^2 = -B[6m^2-12m(1-m)B+4(1-m)(2-m)B^2]\beta^2\,, 
\nonumber \\
&&[m-(1-m)B] \omega_2 = [(5-m)m+(1-m^2)B]\beta^2 \,, \nonumber \\
&&[m-(1-m)B] g_{22} A^2 = 6[m+2m B-(1-m)B^2]\beta^2 \,, \nonumber \\
&&[m-(1-m)B] g_{21} D^2 = -2B [m-2(2-m)B-2(1-m)B^2]\beta^2\,.
\eea

On using the identities (\ref{7}) and (\ref{10}), the coupled 
solution (\ref{2.60}) can be re-expressed as 
\be\label{2.62}
u(x,t) = e^{i\omega_1 t} \frac{A \dn(\Delta,m)}{2}
[\dn(\beta x+\Delta,m)+\dn(\beta x-\Delta,m)]\,, 
\ee
\be\label{2.62a}
v(x,t) = e^{i\omega_2 t} \frac{D \dn^2(\Delta,m)}{2\sn(\Delta,m)}
[\sn(\beta x+\Delta,m)-\sn(\beta x-\Delta,m)]\,,
\ee
where $A, D$ are given by Eq. (\ref{2.61}) while 
$B = {m\sn^2(\Delta,m)}/{\dn^2(\Delta,m)}$. 

{\bf Hyperbolic Limit}

In the limit $m = 1$, the two solutions XIV and XV go over to the
hyperbolic solution III, i.e. in this limit both the solutions can 
be re-expressed as superposition of one pulse   
(i.e. $\sech(\beta x+\Delta) + \sech(\beta x-\Delta)$) and 
one kink (i.e. $\tanh(\beta x +\Delta) - \tanh(\beta x-\Delta)$) solution.

{\bf Solution XVI}

It is easy to check that
\be\label{2.40}
u(x,t) = e^{i\omega_1 t} \frac{A\cn(\beta x,m)\sn(\beta x,m)}
{1+B\cn^2(\beta x,m)}\,,
~~v(x,t) = e^{i\omega_2 t} \frac{D \cn(\beta x,m)\dn(\beta x,m)}
{1+B\cn^2(\beta x,m)}\,,
\ee
where $B > 0$, is an exact solution of the coupled 
Eqs. (\ref{1}) and (\ref{2}) provided
\bea\label{2.41}
&&\omega_1 = [(5m-4)-6(1-m)B]\beta^2\,, \nonumber \\
&&g_{12} D^2 = 6(B+1)[m-2(1-m) B +(1-m)B^2]\beta^2\,, 
\nonumber \\
&&g_{11} A^2 = -2[3m^2 +(9m-5)mB-(1-m)(9m-2)B^2 +3(1-m)^2 B^3]\beta^2\,,
\nonumber \\
&&\omega_2 = [((5m-1)-6(1-m)B]\beta^2\,,~~
g_{22} D^2 = 2m(B+1)[3+2(3m-2)B -3(1-m)^2]\beta^2 \,, \nonumber \\
&&g_{21} A^2  = -6[m^2+m(3m-1)B-3m(1-m) B^2+(1-m)^3 B^3]\beta^2\,.
\eea

On using the identities (\ref{7}) and (\ref{11}), the coupled 
solution (\ref{2.40}) can be re-expressed as 
\be\label{2.42}
u(x,t) = e^{i\omega_1 t} \frac{A \dn^2(\Delta,m)}{2m \cn(\Delta,m)
\sn(\Delta,m)}[\dn(\beta x+\Delta,m)-\dn(\beta x-\Delta,m)]\,,
\ee
\be\label{2.42a}
v(x,t) = e^{i\omega_2 t} \frac{D \dn^2(\Delta,m)}{2\sn(\Delta,m)}
[\sn(\beta x+\Delta,m)-\sn(\beta x-\Delta,m)]\,,
\ee
where $A, D$ are given by Eq. (\ref{2.41}) while
$B = {m\sn^2(\Delta,m)}/{\dn^2(\Delta,m)}$. 

{\bf Solution XVII}

It is easy to check that
\be\label{2.46}
u(x,t) = e^{i\omega_1 t} \frac{A\cn(\beta x,m)\dn(\beta x,m)}
{1+B\cn^2(\beta x,m)}\,,
~~v(x,t) = e^{i\omega_2 t} \frac{D \sn(\beta x,m) \dn(\beta x,m)}
{1+B\cn^2(\beta x,m)}\,,
\ee
where $B > 0$, is an exact solution of the coupled 
Eqs. (\ref{1}) and (\ref{2}) provided
\bea\label{2.47}
&&[m-(1-m)B] \omega_1 = [(5-m)m +(1-m^2)B]\beta^2\,, \nonumber \\
&&[m-(1-m)B] \omega_2 = [(1-m)(1+4m)B+m(5-4m)]\beta^2\,, \nonumber \\ 
&&[m- (1-m)B] g_{12} D^2 = -6 [m +2mB-(1-m)B^2]\beta^2\,,
~~0 < m < 1\,, \nonumber \\
&&[m-(1-m)B] g_{11} A^2 = [6m +2(1+4m)B+2(1+m)m B^2+(1-m)B^3]\beta^2\,, 
\nonumber \\
&&[m-(1-m)B] g_{22} D^2 = -2[3m +4m B -(1-m) B^2]\beta^2\,,
~~[m-(1-m)B] g_{21}  A^2 \nonumber \\
&&= [6m+6mB-2(1-m)(8m-1)B^2-4(1-m)(2m-1)B^3]\beta^2\,.
\eea

On using the identities (\ref{7}) and (\ref{9}), the coupled 
solution (\ref{2.46}) can be re-expressed as 
\be\label{2.48}
u(x,t) = e^{i\omega_1 t} \frac{A\dn^2(\Delta,m)}{2\sn(\Delta,m)}
[\sn(\beta x+\Delta,m)-\sn(\beta x-\Delta,m)]\,, 
\ee
\be\label{2.48a}
v(x,t) = e^{i\omega_2 t} \frac{D \dn(\Delta,m)}{2\sn(\Delta,m)}
[\cn(\beta x-\Delta,m)-\cn(\beta x+\Delta,m)]\,,
\ee
where $A, D$ are given by Eq. (\ref{2.47}) while 
$B = {m\sn^2(\Delta,m)}/{\dn^2(\Delta,m)}$.

{\bf Hyperbolic Limit}

In the limit $m = 1$, the two solutions XVI and XVII go over to the
hyperbolic solution II, i.e. in this limit both the solutions can 
be re-expressed as superposition of one pulse   
(i.e. $\sech(\beta x-\Delta) - \sech(\beta x+\Delta)$) and 
one kink (i.e. $\tanh(\beta x +\Delta) - \tanh(\beta x-\Delta)$) solution.

{\bf Solution XVIII}

It is straightforward to check that
\be\label{2.43}
u(x,t) = e^{i\omega_1 t} \frac{A\sn(\beta x,m)}{1+B\cn^2(\beta x,m)}\,,
~~v(x,t) = e^{i\omega_2 t} \frac{D \sn(\beta x,m) \dn(\beta x,m)}
{1+B\cn^2(\beta x,m)}\,,
\ee
where $B > 0$, is an exact solution of the coupled 
Eqs. (\ref{1}) and (\ref{2}) provided
\bea\label{2.44}
&&(1+B) \omega_1 = [(5-m)B -(1+m)]\beta^2\,,~~m (1+B) g_{12} D^2 
= -6B [m +2mB-(1-m)B^2]\beta^2\,, 
\nonumber \\
&&m(1+B) g_{11} A^2 = 2[m^2 -m(m+1)B+5m(1-m)B^2 -3(1-m)^2 B^3]\beta^2\,, 
\nonumber \\
&&(1+B) \omega_2 = [(5-4m)B-(4m+1)]\beta^2\,,~~m (1+B) g_{22} D^2 
= 2[m +4m B -3(1-m) B^2] B \beta^2\,,
 \nonumber \\
&&m(1+B) g_{21} A^2  = 6[m^2-m(1-m)B +m(1-m)B^2-(1-m)^2 B^3]\beta^2\,.
\eea

On using the identities (\ref{6}) and (\ref{9}), the coupled 
solution (\ref{2.43}) can be re-expressed as 
\be\label{2.45}
u(x,t) = e^{i\omega_1 t} \frac{A \dn(\Delta,m)}{2\cn(\Delta,m)}
[\sn(\beta x+\Delta,m)+\sn(\beta x-\Delta,m)]\,, 
\ee
\be\label{2.45a}
v(x,t) = e^{i\omega_2 t} \frac{D \dn(\Delta,m)}{2\sn(\Delta,m)}
[\cn(\beta x-\Delta,m)-\cn(\beta x+\Delta,m)]\,,
\ee
where $A, D$ are given by Eq. (\ref{2.44}) while 
$B = {m\sn^2(\Delta,m)}/{\dn^2(\Delta,m)}$. 

{\bf Solution XIX}

It is easy to check that
\be\label{2.57}
u(x,t) = e^{i\omega_1 t} \frac{A\sn(\beta x,m)}{1+B\cn^2(\beta x,m)}\,,
~~v(x,t) = e^{i\omega_2 t} \frac{D \sn(\beta x,m) \cn(\beta x,m)}
{1+B\cn^2(\beta x,m)}\,,
\ee
where $B > 0$ is an exact solution of the coupled 
Eqs. (\ref{1}) and (\ref{2}) 
provided
\bea\label{2.58}
&&(1+B) \omega_1 = [(5-m)B-(1+m)]\beta^2\,, \nonumber \\
&&(1+B) g_{12} D^2 = -6B [m+2mB-(1-m)B^2]\beta^2\,, \nonumber \\
&&(1+B) g_{11} A^2 = 2[m-2(2-m)B-(1-m)B^2] \beta^2\,, \nonumber \\
&&(1+B) g_{21} A^2 = 6[m-2(1-m)B-(1-m)B^2]\beta^2 \,, \nonumber \\
&&(1+B) \omega_2 = [(2-m)B-(4+m)]\beta^2 \,, \nonumber \\
&&(1+B) g_{22} D^2 = -2B [m+2(1+m)B-(1-m)B^2] \beta^2\,.
\eea

On using the identities (\ref{6}) and (\ref{11}), the coupled 
solution (\ref{2.57}) can be re-expressed as 
\be\label{2.59}
u(x,t) = e^{i\omega_1 t} \frac{A \dn(\Delta,m)}{2 \cn(\Delta,m)}
[\sn(\beta x+\Delta,m)+\sn(\beta x-\Delta,m)]\,,
\ee
\be\label{2.59a}
v(x,t) = e^{i\omega_2 t} \frac{D \dn^2(\Delta,m)}{2 m\cn(\Delta,m)
\sn(\Delta,m)} [\dn(\beta x-\Delta,m)-\dn(\beta x+\Delta,m)]\,,
\ee
where $A$, $D$ are as given by Eq. (\ref{2.58}) while 
$B = {m\sn^2(\Delta,m)}/{\dn^2(\Delta,m)}$. 

{\bf Hyperbolic Limit}

In the limit $m = 1$, the two solutions XVIII and XIX go over to the
hyperbolic solution IV, i.e. in this limit both the solutions can 
be re-expressed as superposition of one pulse   
(i.e. $\sech(\beta x-\Delta) - \sech(\beta x+\Delta)$) and 
one kink (i.e. $\tanh(\beta x +\Delta) + \tanh(\beta x-\Delta)$) solution.

{\bf Solution XX}

It is strightforward to check that
\be\label{2.50}
u(x,t) = e^{i\omega_1 t} \frac{A\sn(\beta x,m)}{1+B\cn^2(\beta x,m)}\,,
~~v(x,t) = e^{i\omega_2 t} \frac{D \cn(\beta x,m) \dn(\beta x,m)}
{1+B\cn^2(\beta x,m)}\,,~~B > 0\,,
\ee
is an exact solution of the coupled Eqs. (\ref{1}) and (\ref{2}) 
provided
\bea\label{2.51}
&&\omega_1 = \omega_2 = -(1+m)\beta^2\,,~~g_{12} D^2 = 3g_{22} D^2 
= -6B(B+1)\beta^2\,, \nonumber \\
&&g_{21} A^2 = 3 g_{11} A^2 = 6[m-(1-m)B] \beta^2\,.
\eea

On using the identities (\ref{6}) and (\ref{7}), the coupled 
solution (\ref{2.50}) can be re-expressed as 
\be\label{2.53}
u(x,t) = e^{i\omega_1 t} \frac{\sqrt{m}\beta}{\sqrt{2 g_{11}}}
[\sn(\beta x+\Delta,m)+\sn(\beta x-\Delta,m)]\,, 
\ee
\be\label{2.53a}
v(x,t) = e^{i\omega_2 t} \frac{\sqrt{m}\beta}
{\sqrt{2|g_{22}|}}[\sn(\beta x+\Delta,m)-\sn(\beta x-\Delta,m)]\,,
\ee
where $B = {m\sn^2(\Delta,m)}/{\dn^2(\Delta,m)}$. 

{\bf Hyperbolic Limit}

In the limit $m = 1$, the solution XX goes over to the
hyperbolic solution I, i.e. in this limit the solution can 
be re-expressed as superposition of two kink solutions   
(i.e. $\tanh(\beta x+\Delta) \pm \tanh(\beta x-\Delta)$).

{\bf Solution XXI}

It is easy to check that
\be\label{2.19}
u(x,t) = e^{i\omega_1 t} \frac{A\cn(\beta x,m)}{1+B\cn^2(\beta x,m)}\,,
~~v(x,t) = e^{i\omega_2 t} \frac{D\dn(\beta x,m)}{1+B\cn^2(\beta x,m)}\,,
\ee
where $B > 0$, is an exact solution of the coupled 
Eqs. (\ref{1}) and (\ref{2}) provided
\bea\label{2.20}
&&\omega_1 = [(2m-1) +\frac{6m}{B}]\beta^2\,,~~(1-m) g_{12} D^2 
= -6[(1-m)B]+\frac{m}{B}]\beta^2\,, \nonumber \\
&&g_{11} A^2 = 2 \bigg [(1-m)^2 B^2-(4m-3)(1-m)B-7m(1-m)B+\frac{3m^2}{B}
\bigg ]\beta^2 \,, \nonumber \\
&&\omega_2 = [(5m-4)+\frac{6m}{B}]\beta^2\,,~~(1-m) g_{22} D^2 = 
2[2(1-m)-(1-m)B-\frac{3m}{B}]\beta^2\,, \nonumber \\
&&(1-m) g_{21} A^2 = \frac{6\beta^2}{B} [m-(1-m)B] [1-(1-m)(B+1)^2]\,.
\eea
Note that this solution is only valid if $0 < m < 1$.

On using the identities (\ref{8}) and (\ref{10}), the coupled 
solution (\ref{2.19}) can be re-expressed as 
\be\label{2.21}
u(x,t) = e^{i\omega_1 t} \frac{A\dn^2(\Delta,m)}{2\cn(\Delta,m)}
[\cn(\beta x+\Delta,m)+\cn(\beta x-\Delta,m)]\,,
\ee
\be\label{2.21a}
v(x,t) = e^{i\omega_2 t} \frac{D\dn(\Delta,m)}{2}
[\dn(\beta x+\Delta,m)+\dn(\beta x-\Delta,m)]\,,
\ee
where $A, D$ are as given by Eq. (\ref{2.20}) while 
$B = {m\sn^2(\Delta,m)}/{\dn^2(\Delta,m)}$. 

{\bf Solution XXII}

It is straightforward to check that
\be\label{2.25}
u(x,t) = e^{i\omega_1 t} \frac{A\cn(\beta x,m) \sn(\beta x,m)}
{1+B\cn^2(\beta x,m)}\,,
~~v(x,t) = e^{i\omega_2 t} \frac{D \sn(\beta x,m)\dn(\beta x,m)}
{1+B\cn^2(\beta x,m)}\,,
\ee
where $B > 0$, is an exact solution of the coupled 
Eqs. (\ref{1}) and (\ref{2}) provided
\bea\label{2.26}
&&(B+1) \omega_1 = [(2-m)B-(4+m)]\beta^2\,,~~(1-m)(B+1) g_{11} A^2 
\nonumber \\
&&= -[(1-m)(2-m)B^3+(1-m)(5m-4)B^2 +12m(1-m)B-6m^2]\beta^2 \,, \nonumber \\
&&(1-m)(B+1) g_{12} D^2 = -6[1- (1-m)(1+B)^2]\beta^2\,, \nonumber \\
&&(B+1) \omega_2 = [(5-4m)B-(4m+1)]\beta^2\,, \nonumber \\
&&(1-m) (B+1) g_{21} A^2 = [1-15m+20m^2-(1-m)(5+2m)B \nonumber \\
&&+(1-m)(2m-1)B^2 +(1-m)(2m-1)B^3]\beta^2\,, \nonumber \\
&&(1-m)(B+1) g_{22} D^2 = -2[3-8(1-m)B -(1-m) B^2]\beta^2\,.
\eea
Note that this solution is only valid for $0 < m < 1$.

On using the identities (\ref{9}) and (\ref{11}), the coupled 
solution (\ref{2.25}) can be re-expressed as 
\be\label{2.27}
u(x,t) = e^{i\omega_1 t} \frac{A\dn^2(\Delta,m)}{2m \cn(\Delta,m)
\sn(\Delta,m)}[\dn(\beta x-\Delta,m)-\dn(\beta x+\Delta,m)]\,, 
\ee
\be\label{2.27a}
v(x,t) = e^{i\omega_2 t} \frac{D\dn(\Delta,m)}{2\sn(\Delta,m)} 
[\cn(\beta x-\Delta,m)-\cn(\beta x-\Delta,m)]\,,
\ee
where $A, D$ are as given by Eq. (\ref{2.26}) while 
$B = {m\sn^2(\Delta,m)}/{\dn^2(\Delta,m)}$.

\section{Superposed Solutions of a Coupled Nonlocal mKdV Model}

Let us consider the following model of coupled nonlocal mKdV
equations
\be\label{4.1}
u_t(x,t)+u_{xxx}+6[g_{11} u(x,t) u(-x,-t)+g_{12} v(x,t) v(-x,-t)]
u_x(x,t) = 0\,,
\ee
\be\label{4.2}
v_t(x,t)+v_{xxx}+6[g_{21} u(x,t) u(-x,-t)+g_{22} v(x,t) v(-x,-t)]
v_x(x,t) = 0\,.
\ee

We now show that these coupled nonlocal equations admit a large number
of exact solutions. In particular, there are two types of solutions to
these coupled equations depending on if $v(x,t)$ is proportional to 
$u(x,t)$ or not which we discuss one by one. In this section we only
discuss those solutions where the two coupled fields $u$ and $v$ are
distinct (and not proportional to each other) while in Appendix B
we discuss those solutions of the nonlocal coupled mKdV model in
which the two fields $v$ and $u$ are proportional to each other. 

One major difference between the local and the nonlocal case is that, 
unlike the local case, the 
solutions of the nonlocal mKdV Eqs. (\ref{4.1}) and (\ref{4.2}) 
 are not invariant with respect to shifts in $x$ and $t$. 

\subsection{Solutions When $u(x,t)$ and $v(x,t)$ are Distinct}

It turns out that in case $v(x,t)$ and $u(x,t)$ are distinct then 
the coupled Eqs. (\ref{4.1}) and (\ref{4.2}) admit a number of exact 
solutions which we now present one by one.

{\bf Solution I}

It is easy to show that 
\be\label{5.1}
u(x,t) = A \dn(\xi,m)\,,~~v(x,t) = B \sqrt{m} \sn(\xi,m)\,,~~
\xi = \beta(x-ct)\,,
\ee
is an exact solution to the coupled Eqs. (\ref{4.1}) and (\ref{4.2})
provided
\be\label{5.2}
g_{11} A^2 + g_{12} B^2 = g_{21} A^2 + g_{22} B^2 = \beta^2\,,
\ee
\be\label{5.3}
c = (2-m)\beta^2 - 6g_{12} B^2\,, 
\ee
\be\label{5.4}
2 (g_{22} - g_{12}) B^2 = \beta^2\,.
\ee

{\bf Solution II}

It is straightforward to show that 
\be\label{5.5}
u(x,t) = A \sqrt{m} \cn(\xi,m)\,,~~v(x,t) = B \sqrt{m} \sn(\xi,m)\,,~~
\xi = \beta(x-ct)\,,
\ee
is an exact solution to the coupled Eqs. (\ref{4.1}) and (\ref{4.2})
provided Eqs. (\ref{5.2}) and (\ref{5.4}) are satisfied and if further
\be\label{5.6}
c = (5-4m)\beta^2 - 6 [g_{12} B^2 +(1-m) g_{11} A^2]\,.
\ee

In the limit $m = 1$, both the solutions XXIII and XXIV go over to
the hyperbolic solution
\be\label{5.7}
u(x,t) = A \sech(\xi)\,,~~v(x,t) = B \tanh(\xi)\,,~~
\xi = \beta(x-ct)\,,
\ee
provided Eqs. (\ref{5.2}) and (\ref{5.4}) are satisfied and if further
\be\label{5.8}
c = \beta^2 - 6 [g_{12} B^2 +(1-m) g_{11} A^2]\,.
\ee

{\bf Solution III}

It is easy to show that 
\be\label{5.9}
u(x,t) = A \dn(\xi,m)\,,~~v(x,t) = B \sqrt{m} \cn(\xi,m)\,,~~
\xi = \beta(x-ct)\,,~~ 0 < m < 1\,,
\ee
is an exact solution to the coupled Eqs. (\ref{4.1}) and (\ref{4.2})
provided Eqs. (\ref{5.2}) and (\ref{5.4}) are satisfied and if further
\be\label{5.10}
c = (5-4m)\beta^2 - 6 [g_{12} B^2 +(1-m) g_{11} A^2]\,.
\ee
Note that if $m = 1$, then $v(x,t) = u(x,t)$ and that solution has been
discussed in Appendix B.

{\bf Solution IV}

It is straightforward to show that 
\be\label{5.11}
u(x,t) = \frac{A}{\dn(\xi,m)}\,,~~v(x,t) = B \sqrt{m} 
\frac{\sn(\xi,m)}{\dn(\xi,m)}\,,~~\xi = \beta(x-ct)\,,
\ee
is an exact solution to the coupled Eqs. (\ref{4.1}) and (\ref{4.2})
provided
\be\label{5.12}
g_{11} A^2 - g_{12} B^2 = g_{21} A^2 - g_{22} B^2 = (1-m) \beta^2\,,
\ee
\be\label{5.13}
c = (2-m)\beta^2 + 6 g_{12} B^2\,, 
\ee
\be\label{5.14}
2 (g_{22} - g_{12}) B^2 = (1-m) \beta^2\,.
\ee

{\bf Solution V}

It is easy to show that 
\be\label{5.15}
u(x,t) = \frac{A}{\dn(\xi,m)}\,,~~v(x,t) 
= \frac{B\sqrt{m} \cn(\xi,m)}{\dn(\xi,m)}\,,~~\xi = \beta(x-ct)\,,
\ee
is an exact solution to the coupled Eqs. (\ref{4.1}) and (\ref{4.2})
provided Eq. (\ref{5.13}) is satisfied and further
\be\label{5.16}
g_{11} A^2 - (1-m) g_{12} B^2 = g_{21} A^2 - (1-m) g_{22} B^2 
= (1-m) \beta^2\,,
\ee
\be\label{5.17}
2 (g_{22} - g_{12}) B^2 = -\beta^2\,.
\ee

{\bf Solution VI}

It is straightforward to show that 
\be\label{5.18}
u(x,t) = \frac{A\sqrt{m} \sn(\xi,m)}{\dn(\xi,m)}\,,~~v(x,t) 
= \frac{B\sqrt{m} \cn(\xi,m)}{\dn(\xi,m)}\,,~~\xi = \beta(x-ct)\,,
\ee
is an exact solution to the coupled Eqs. (\ref{4.1}) and (\ref{4.2})
provided 
\be\label{5.19}
g_{11} A^2 + (1-m) g_{12} B^2 = g_{21} A^2 + (1-m) g_{22} B^2 
= -(1-m) \beta^2\,,
\ee
\be\label{5.20}
c = (5-4m)\beta^2 +6[g_{11}A^2+g_{12}B^2]\,,
\ee
\be\label{5.21}
2(g_{11}-g_{21})A^2 -2 (g_{12} - g_{22}) B^2 = m\beta^2\,.
\ee

{\bf Solution VII}

The coupled nonlocal MKdV Eqs. (\ref{4.1}) and (\ref{4.2})
admit the solution
\be\label{5.22}
u(x,t) = 1- \frac{A}{B+\cosh^2(\xi)}\,,~~ v(x,t) 
= \frac{\alpha A}{B+\cosh^2(\xi)}\,, ~~~A, B > 0\,,
\ee
provided
\bea\label{5.23}
&&c = 4\beta^2 + 6 g_{11}\,,~~g_{21} = g_{11} < 0\,,~~g_{22} = g_{12}\,,
~~A g_{11} = -(2B+1)\beta^2\,, \nonumber \\
&&(g_{11}+\alpha^2 g_{12})A^2 = -4B(B+1) \beta^2\,.
\eea

On comparing it with the identity (\ref{7}), we can re-express 
solution (\ref{5.22}) as
\be\label{5.24}
u(x,t) = 1- \frac{\beta}{\sqrt{|g_{11}+\alpha^2 g_{22}|}}
[\tanh(\xi+\Delta) -\tanh(\xi-\Delta)]\,,
\ee
and 
\be\label{5.25}
v(x,t)  = \frac{\alpha \beta}{\sqrt{|g_{11}+\alpha^2 g_{22}|}}
[\tanh(\xi+\Delta) -\tanh(\xi-\Delta)]\,,
\ee
where $B = \sinh^2(\Delta)$.

{\bf Solution VIII}

Yet another hyperbolic superposed solution to the coupled Eqs. 
(\ref{4.1}) and (\ref{4.2}) is 
\be\label{5.26}
u(x,t) = 1- \frac{A}{B+\cosh^2(\xi)}\,,~~v(x,t)
= -b -\frac{\alpha A}{B+\cosh^2(\xi)}\,,~~A, B, b > 0
\ee
provided
\bea\label{5.27}
&&c-4\beta^2 = g_{12} + b^2 g_{22} = g_{11}+b^2 g_{12}\,,
\nonumber \\
&&A(g_{11} - g_{12} \alpha b) = = A(g_{21}-g_{22} \alpha b)
= -(2B+1)\beta^2\,, \nonumber \\
&&(g_{11}+\alpha^2 g_{12}) A^2 = 
= (g_{21}+\alpha^2 g_{22}) = -4B(B+1) \beta^2\,. 
\eea

On comparing it with the identity (\ref{7}), we can re-express 
solution (\ref{5.26}) as
\be\label{5.28}
u(x,t)  = 1- \frac{\beta}{\sqrt{|g_{11}+\alpha^2 g_{22}|}}
[\tanh(\xi+\Delta) -\tanh(\xi-\Delta)]\,,
\ee
and 
\be\label{5.29}
v(x,t) = -b - \frac{\alpha \beta}{\sqrt{|g_{11}+\alpha^2 g_{22}|}}
[\tanh(\xi+\Delta) -\tanh(\xi-\Delta)]\,,
\ee
where $B = \sinh^2(\Delta)$.

\section{Conclusion and Open Problems}

In this paper we have partially extended the notion of superposition to
the coupled nonlocal equations. In particular, we showed that the
Ablowitz-Musslimani variant of the coupled nonlocal NLS equations as well as the 
coupled nonlocal mKdV equations admit solutions which can be re-expressed
as the superposition of two hyperbolic as well as two periodic kink
and pulse solutions. In particular, in Sec. II, for the 
Ablowitz-Musslimani NLS case we obtained seven hyperbolic superposed 
solutions as well as fifteen periodic superposed solutions. On the 
other hand, for the nonlocal mKdV case, in Sec. III  we obtained two
superposed hyperbolic solutions. Besides, in case the two coupled
fields are proportional to each other then in the Ablowitz-Musslimani
case we obtained four superposed periodic solutions while in the mKdV
case we obtained four superposed periodic and one superposed hyperbolic
superposed solution. In addition, for completeness we have also presented
some other solutions admitted by these models. 

This paper raises several important questions which are still not understood. 
Someof these questions are

\begin{enumerate}

\item Unlike the coupled nonlocal Ablowitz-Musslimani case, we have not
been able to obtain superposed periodic solutions in the nonlocal mKdV
case when the two fields are not proportional to each other. It would
be worthwhile looking for such solutions.  

\item For the nonlocal mKdV case, in the hyperbolic case we have only been 
able to obtain solutions which can be re-expressed as a superposition
of a kink and an antikink solution but not as a superposition of two 
kink or two pulse solutions. It is clearly of interest to obtain
such solutions.

\item It is clearly of interest to discover similar superposed solutions
in the case of other nonlocal equations such as the nonlocal KdV equation, the 
nonlocal Hirota equation, etc. As a first step in that direction one needs to construct
models with coupled nonlocal KdV and coupled nonlocal Hirota equations.

\item Finally, what is the interpretation of such superposed solutions?
Do they correspond to bound state or merely an excitation of two kink
or two pulse solutions? 

Hopefully, one can find answer to some of the questions raised above.

\end{enumerate}

{\bf Acknowledgment}

One of us (AK) is grateful to Indian National Science Academy (INSA) for the
award of INSA Senior Scientist position at Savitribai Phule Pune University. 
The work at Los Alamos National Laboratory was carried out under the auspices 
of the U.S. DOE and NNSA under Contract No. DEAC52-06NA25396. 

{\bf Appendix A: Solutions of a Coupled Ablowitz-Musslimani nonlocal 
NLS Model when $v(x,t) \propto u(x,t)$}

In case $v(x,t) = \alpha u(x,t)$ where $\alpha$ is a number, in that case the
two Eqs. (\ref{1}) and (\ref{2}) are consistent with each other 
provided
\be\label{3.1}
g_{11}+\alpha^2 g_{12} = g_{21}+\alpha^2 g_{22}\,,
\ee
and we only need to solve the single equation
\be\label{3.2}
iu_{x,t}+u_{xx}(x)+ g u^2(x,t)u^{*}(-x,t) = 0\,,~~g = g_{11}+
\alpha^2 g_{12}\,,
\ee
which is effectively the uncoupled nonlocal Ablowitz-Musslimani variant of 
the NLS for which we have \cite{ks14} already obtained a large number of
solutions. However, it turns out that we have missed several solutions 
which we now present one by one.

As mentioned in Sec. II, one major difference between the local and 
the nonlocal case is that, unlike the local case, the 
solutions of the nonlocal mKdV Eq. (\ref{3.2})  
 are not invariant with respect to shift in $x$.

{\bf Solution I}

In \cite{ks14} we had shown that $e^{i\omega t} \dn(\beta x,m)$ as well as 
$e^{i\omega t} \dn[\beta(x+K[m]),m]$ are the exact solutions of the nonlocal 
Eq. (\ref{3.2}). 
Remarkably, it turns out that not only $e^{i\omega t} \dn[\beta x,m]$ and 
$e^{i\omega t} \dn[\beta(x+K[m]),m]$ but even their superposition is a 
solution of the nonlocal NLS Eq. (\ref{3.2}). In particular, it is 
straightforward to check that
\be\label{3.3}
u(x,t) = e^{i\omega t} [A\dn(\beta x,m)+ \frac{B\sqrt{1-m}}
{\dn(\beta x,m)}]\,,
\ee
is also an exact solution of the nonlocal Eq. (\ref{3.2}) provided
\be\label{3.4}
g > 0\,,~~ g A^2 = \beta^2\,,~~B = \pm A\,,~~
\omega = [2-m \pm 6\sqrt{1-m}]\beta^2\,,
\ee
where $g$ is as given by Eq. (\ref{3.2}) while the $\pm$ sign in 
$B = \pm A$ and in $\omega$ are correlated.

{\bf Solution II: Peregrine Soliton}

It is easy to check that the celebrated Peregrine soliton solution 
\cite{per83,dj} of the local NLS is also a solution of the nonlocal 
Eq. (\ref{3.2}), i.e. in particular, it is straightforward to check that 
\be\label{3.5}
u(x,t) = \frac{1}{\sqrt{2g}} \bigg [1-\frac{4(1+2it)}
{(1+2x^2+4t^2)} \bigg ] e^{it}\,,~~g > 0\,,
\ee
is an exact solution of Eq. (\ref{3.2}).

{\bf Solution III: Akhmedive-Eleonskii-Kulagin Breather Solution}

It is easy to check that even the celebrated Akhmediev-Eleonskii-Kulagin 
breather solution \cite{dj,akh85} of the local NLS, i.e. 
\be\label{3.6}
u(x,t) = \sqrt{\frac{a^2}{2g}} e^{ia^2 t} \bigg 
[\frac{b^2 \cosh(\theta) +ib\sqrt{2-b^2}}{\sqrt{2}\cosh(\theta)-\sqrt{2-b^2}
\cos(abx)} \bigg ]\,,~~g > 0\,,
\ee
where $\theta = a^2 b \sqrt{2-b^2} t$, is also a solution of 
the nonlocal Eq. (\ref{3.2}). 

{\bf Solution IV: Kuznetsov-Ma Soliton}

Remarkably, even the celebrated Kuznetsov-Ma soliton solution 
\cite{dj,kuz77,ma79} 
of the local NLS is also a solution of the nonlocal Eq. (\ref{3.2}). 
In particular, it is easy to check that
\be\label{3.7}
u(x,t) =\frac{a}{\sqrt{2g}} e^{ia^2 t}\bigg [1
+\frac{2m(m\cos(\theta) +in\sin(\theta)}{n\cosh(\sqrt{2} ma x)
+\cos(\theta)} \bigg ]\,,~~g > 0\,,
\ee
where $n^2 = 1+m^2\,,~~\theta = 2m n a^2 t$, is an exact solution of the 
nonlocal Eq. (\ref{3.2}). 

We now show that unlike the local NLS or the Yang version of the nonlocal NLS 
\cite{yang18}, the Ablowitz-Musslimani variant of the
nonlocal NLS admits complex PT-invariant periodic and hyperbolic 
superposed solutions. 

{\bf Solution V}

It is straightforward to show that the nonlocal Eq. (\ref{3.2}) 
admits the complex PT-invariant periodic solution
\be\label{3.8}
u(x,t) = e^{i\omega t}[A\dn(\beta x,m) + iB\sqrt{m} \sn(\beta x,m)]\,,
\ee
provided
\be\label{3.9}
B = \pm A\,,~~A = \frac{\beta}{2g}\,,~~\omega = 
-\frac{(2m-1)}{2}\beta^2\,,~~g > 0\,.
\ee

{\bf Solution VI}

Another complex PT-invariant periodic solution of the nonlocal 
Eq. (\ref{3.2}) is
\be\label{3.10}
u(x,t) = e^{i\omega t}[A\sqrt{m} \cn(\beta x,m) 
+ iB\sqrt{m} \sn(\beta x,m)]\,,
\ee
provided
\be\label{3.11}
B = \pm A\,,~~g >0\,,~~A = \frac{\beta}{2g}\,,~~\omega = 
-\frac{(2-m)}{2}\beta^2\,.
\ee

{\bf Solution VII}

In the limit $m = 1$, both the solutions V and VI go over to the 
complex PT-invariant hyperbolic solution 
\be\label{3.12}
u(x,t) = e^{i\omega t}[A\sech(\beta x) 
+ iB \tanh(\beta x)]\,,
\ee
provided
\be\label{3.13}
B = \pm A\,,~~g > 0\,,~~A = \frac{\beta}{2g}\,,~~\omega = 
-\frac{\beta^2}{2}\,.
\ee

{\bf Solution VIII}

Remarkably, the nonlocal Eq. (\ref{3.2}) also admits the
complex PT-invariant periodic solution
\be\label{3.14}
u(x,t) = e^{i\omega t}[A\sqrt{m} \sn(\beta x,m) + iB \dn(\beta x,m)]\,,
\ee
provided the {\it same} relations as given in Eq. (\ref{3.9}) are satisfied. 

{\bf Solution IX}

Another complex PT-invariant periodic solution of the nonlocal 
Eq. (\ref{1}) is
\be\label{3.15}
u(x,t) = e^{i\omega t}[A\sqrt{m} \sn(\beta x,m) 
+ iB\sqrt{m} \cn(\beta x,m)]\,,
\ee
provided the {\it same} relations as in Eq. (\ref{3.11}) are satisfied.

{\bf Solution X}

In the limit $m = 1$, both the solutions VIII and IX go over to the 
complex PT-invariant hyperbolic solution 
\be\label{3.16}
u(x,t) = e^{i\omega t}[A\tanh(\beta x) + iB \sech(\beta x)]\,,
\ee
provided the {\it same} relations as in Eq. (\ref{3.13}) are satisfied.
Thus the Ablowitz-Musslimani variant of the nonlocal NLS is rather unusual in
the sense that the same model (i.e. with $g > 0$) admits not only the
kink and the pulse solutions but also the complex PT-invariant pulse and
kink solutions with the PT-eigenvalue +$1$ as well as $-1$.

We now show that Eq. (\ref{3.2}) also satisfies four novel periodic solutions 
which can be re-expressed as the superposition of either the two periodic 
kink or the two periodic pulse solutions $\sn(x,m)$ or $\dn(x,m)$, respectively.

{\bf Solution XI}

It is easy to check that the nonlocal NLS Eq. (\ref{3.2}) 
admits the periodic solution
\be\label{3.17}
u(x,t) = e^{i\omega t} [\frac{A \dn(\beta x,m) 
\cn(\beta x,m)}{1+B\cn^2(\beta x,m)}]\,,
~~B > 0\,,
\ee
provided $g < 0$ and further 
\bea\label{3.18}
&&0 < m < 1\,,~~B = +\frac{\sqrt{m}}{1-\sqrt{m}}\,,
\nonumber \\
&&\omega = -[1+m+6\sqrt{m}]\beta^2 < 0\,,~~g < 0\,,~~
|g| A^2 = \frac{4 \sqrt{m} \beta^2}{(1-\sqrt{m})^2}\,.
\eea
Note that this solution is not valid for $m = 1$, i.e. the nonlocal NLS
Eq. (\ref{1}) does not admit a corresponding hyperbolic solution.

On comparing Eqs. (\ref{3.17}) and the identity (\ref{7}) and 
using Eq. (\ref{3.18}), 
one can re-express the periodic solution (\ref{3.17}) 
as the superposition of a periodic kink and a periodic antikink 
solution, i.e.
\be\label{3.19}
u(x,t) = e^{i\omega t} \sqrt{\frac{m}{2|g|}} \beta
 [\sn(\beta x +\Delta, m) -\sn(\beta x- \Delta, m)]\,. 
\ee
Here $\Delta$ is defined by $\sn(\sqrt{m}\Delta,1/m) = \pm m^{1/4}$,
where use has been made of the identity \cite{boyd} 
\be\label{3.20}
\sqrt{m} \sn(y,m) = \sn(\sqrt{m} y,1/m) \,. 
\ee

{\bf Solution XII}

Remarkably, Eq. (\ref{1}) also admits another periodic 
solution
\be\label{3.21}
u(x,t) = e^{i\omega t} \frac{A \sn(\beta x,m)}{1+B\cn^2(\beta x,m)}\,,
~~B > 0\,,
\ee
provided
\bea\label{3.22}
&&0 < m < 1\,,~~B = \frac{\sqrt{m}}{1-\sqrt{m}}\,,~~g < 0\,,
\nonumber \\
&&\omega = [6\sqrt{m}-(1+m)]\beta^2\,,~~|g| A^2 = 4\sqrt{m} \beta^2\,.
\eea 
Note that this solution too is not valid in the hyperbolic limit of 
$m = 1$. On comparing the solution (\ref{3.21}) and the novel identity 
(\ref{6}) and using Eq. (\ref{3.22}), 
the periodic solution XII given by Eq. (\ref{3.21}) can be
re-expressed as the superposition of the two periodic kink solutions
\be\label{3.23}
u(x,t) = i e^{i\omega t} \sqrt{\frac{m}{|g|}} \beta  
\bigg [\sn(\beta x +\Delta, m)+\sn(\beta x -\Delta, m) \bigg ]\,. 
\ee
Here $\Delta$ is defined by $\sn(\sqrt{m}\Delta,1/m) = \pm m^{1/4}$,
where use has been made of the identity (\ref{3.20}).

It is worth noting that for both the solutions XI and XII, not only 
the value of $B$ is the same but even $g < 0$ for both the solutions.

{\bf Solution XIII}

It is easy to check that the nonlocal NLS Eq. (\ref{3.2})
admits another periodic solution
\be\label{3.24}
u(x,t) = e^{i\omega t} \frac{A \sn(\beta x,m) 
\cn(\beta x,m)}{1+B\cn^2(\beta x,m)}\,,
\ee
provided
\bea\label{3.25}
&&0 < m < 1\,,~~B = \frac{1-\sqrt{1-m}}{\sqrt{1-m}}\,,
~~g < 0\,, \nonumber \\
&&\omega = (2-m-6\sqrt{1-m})\beta^2\,,~~ A^2 
= \frac{2(1-\sqrt{1-m})^2 \beta^2}{\sqrt{1-m}}\,.
\eea 
Note that whereas this solution is valid if $g < 0$, the same
solution in the Yang's nonlocal case, as well as in the local NLS case 
is valid only if $g > 0$. 

On comparing the solution (\ref{3.24}) with the identity (\ref{11})
and using Eq. (\ref{3.25}) we find that the solution 
as given by Eq. (\ref{3.24}) can be re-expressed as a 
superposition of the two periodic pulse solutions, i.e.
\be\label{3.26}
u(x,t) = e^{i\omega t}\beta \sqrt{\frac{1}{2|g|}} 
\big (\dn[\beta(x)-K(m)/2,m] - \dn[\beta(x)+K(m)/2,m] \big )\,.
\ee

{\bf Solution XIV}

Remarkably, the nonlocal NLS Eq. (\ref{3.2}) also 
admits another periodic solution
\be\label{3.27}
u(x,t) = e^{i\omega t} 
\frac{A \dn(\beta x,m)}{1+B\cn^2(\beta x,m)}\,,
\ee
provided
\bea\label{3.28}
&&0 < m < 1\,,~~ B = \frac{1-\sqrt{1-m}}{\sqrt{1-m}}\,, 
~~g > 0\,, \nonumber \\
&&\omega = [2-m+6\sqrt{1-m}]\beta^2\,,
~~A^2 = \frac{4}{\sqrt{1-m}} \beta^2\,.
\eea 
Note that this solutions is also not valid for $m = 1$.
Thus for this solution $g > 0$, $\omega > 0$.

On comparing the solution (\ref{3.27}) and the identity (\ref{10}) 
and using Eq. (\ref{3.28}), the periodic solution (\ref{3.27}) can be
re-expressed as a superposition of the two periodic pulse solutions, i.e.
\be\label{3.29}
u(x,t) = e^{i\omega t} \frac{\beta}{\sqrt{g}} 
[\dn(\beta x +K(m)/2, m) +\dn(\beta x -K(m)/2, m)]\,,
\ee
where $\Delta = \pm K(m)/2$ 

It is worth noting that for both the superposed periodic pulse 
solutions XIII and XIV, while the 
value of $B$ is the same but the value of $g$ is opposite
for the two solutions. 

Note that out of the 14 solutions, the solutions XI to XIV are only
valid if $m \ne 1$. Further, while the solutions XI, XII and XIII are 
valid if $g < 0$, the remaining eleven solutions are valid if $g > 0$. 

{\bf Appendix B: Solutions of a Coupled Nonlocal mKdV Model When 
$v(x,t) \propto u(x,t)$}

In case $v(x,t) = \alpha u(x,t)$ where 
$\alpha$ is a number, then the two coupled Eqs. (\ref{4.1}) 
and (\ref{4.2}) are consistent with each other provided
\be\label{4.3}
g_{11}+\alpha^2 g_{12} = g_{21}+\alpha^2 g_{22}\,.
\ee
Remarkably, in this case we only need to solve an uncoupled 
nonlocal mKdV equation 
\be\label{4.4}
u_t(x,t)+u_{xxx}+6g u(x,t) u(-x,-t) u_x(x,t) = 0\,,~~g = 
g_{11}+\alpha^2 g_{12}\,.
\ee
In a recent paper \cite{ks22c} we have obtained several exact solutions 
to an uncoupled nonlocal mKdV equation using which we can immediately 
write down the
exact solutions to the coupled nonlocal mKdV Eqs. (\ref{4.1}) and
(\ref{4.2}) in case the two nonlocal fields $v(x,t)$ and $u(x,t)$
are proportional to each other. 

{\bf Solution I}

It is easy to show that one of the exact solution of the nonlocal 
mKdV Eq. (\ref{4.4}) is
\be\label{4.5}
u(x,t) = A \dn[\beta (x-ct), m] \,,
\ee
provided
\be\label{4.6}
g > 0\,,~~g A^2 = \beta^2\,,~~c = (2-m)\beta^2\,.
\ee

{\bf Solution II}

Similarly 
\be\label{4.7}
u(x,t) = A \sqrt{m} \cn[\beta (x-ct), m] \,,
\ee
is an exact solution to Eq. (\ref{4.4}) provided
\be\label{4.8}
g > 1\,,~~g A^2 = \beta^2\,,~~c = (2m-1)\beta^2\,.
\ee

{\bf Solution III}

Remarkably, even a linear superposition of the solutions I and II is also a
solution of Eq. (\ref{4.4}), i.e.
\be\label{4.9}
u(x,t) = A \dn[\beta (x-ct), m] +B \sqrt{m} \cn[\beta (x-ct), m] 
\big ]\,,
\ee
provided
\be\label{4.10}
g > 0\,,~~ 4g A^2 = \beta^2\,,~~B = \pm A\,,~~c = \frac{(1+m)}{2} \beta^2\,.
\ee

{\bf Solution IV}

In the limit $m = 1$, the solutions I, II and III (with $B = A$) go over to
the hyperbolic solution
\be\label{4.11}
u(x,t) = A \sech[\beta (x-ct)]\,,
\ee
provided
\be\label{4.12}
g > 1\,,~~gA^2 = \beta^2\,,~~v = \beta^2\,,
\ee
while solution III with $B = -A$ goes over to the vacuum solution
$u = 0$.

{\bf Solution V}

Remarkably, unlike the local mKdV, for the nonlocal case even 
\be\label{4.13}
u(x,t) = A \sqrt{m} \sn[\beta (x-ct), m]\,,
\ee
is an exact solution to Eq. (\ref{4.4}) provided
\be\label{4.14}
g > 0\,,~~g A^2 = \beta^2\,,~~c = -(1+m)\beta^2\,.
\ee

{\bf Solution VI}

In the limit $m = 1$, the solution V goes over to the hyperbolic solution
\be\label{4.15}
u(x,t) = A \tanh[\beta (x-ct)] \,,
\ee
provided
\be\label{4.16}
g > 0\,,~~g A^2 = \beta^2\,,~~c = -2\beta^2\,.
\ee
Thus unlike the local mKdV, the nonlocal attractive mKdV Eq. (\ref{4.4}) 
(i.e. with $g = +1$) admits both the kink and the pulse solutions.

Unlike the local case, the 
solutions of the nonlocal mKdV Eq. (\ref{4.4})
 are not invariant with respect to shifts in $x$ and $t$. For example, while
$A \dn[\beta(x-ct+x_0), m]$ is an exact solution of the local mKdV
no matter what $x_0$ is, it is not an exact solution of the nonlocal Eq. 
(\ref{4.4}). However, for the special values of $x_0$, $\sn(x,m)$, $\cn(x,m)$
and $\dn(x,m)$ are still the solutions of the nonlocal Eq. (\ref{4.4}).
In particular, we now show that when $x_0 = K(m)$, where $K(m)$ is the 
complete elliptic integral of the first kind, there are exact solutions of 
the nonlocal Eq. (\ref{4.4}) in both the focusing ($g > 0$) and
the defocusing ($g < 0$) cases. This is because of the relations \cite{as} 
\bea\label{4.17}
&&\dn[x+K(m),m] = \frac{\sqrt{1-m}}{\dn(x,m)}\,,~~\sn[x+K(m),m]
= \frac{\cn(x,m)}{\dn(x,m)}\,, \nonumber \\
&&\cn[x+K(m),m] = -\frac{\sqrt{1-m} \sn(x,m)}{\dn(x,m)}\,.
\eea

{\bf Solution VII}

It is easy to show that 
\be\label{4.18}
u(x,t) = \frac{A}{\dn[\beta (x-ct), m]}\,,
\ee
is an exact solution to the Eq. (\ref{4.4}) provided
\be\label{4.19}
g > 0\,,~~g A^2 = (1-m) \beta^2\,,~~c = (2-m)\beta^2\,.
\ee

{\bf Solution VIII}

It is easy to show that 
\be\label{4.20}
u(x,t) = \frac{A\sqrt{m} \sn[\beta(x-ct), m]}{\dn[\beta (x-ct), m]}\,,
\ee
is an exact solution to the nonlocal mKdV Eq. (\ref{4.4}) provided
\be\label{4.21}
g < 0\,,~~|g| A^2 = (1-m) \beta^2\,,~~c = (2m-1)\beta^2\,.
\ee

{\bf Solution IX}

It is easy to show that 
\be\label{4.22}
u(x,t) = \frac{A\sqrt{m} \cn[\beta(x-ct), m]}{\dn[\beta (x-ct), m]}\,,
\ee
is an exact solution to the Eq. (\ref{4.4}) provided
\be\label{4.23}
g < 0\,,~~|g| A^2 = \beta^2\,,~~c = -(1+m)\beta^2\,,~~m \ne 1\,.
\ee

{\bf Solution X}

Remarkably, it turns out that not only $\dn[\beta(x-ct),m]$ and 
$\dn[\beta(x-ct+K[m]),m]$ but even their superposition is a solution
of the nonlocal mKdV Eq. (\ref{4.4}). In particular, it is easy to check
that
\be\label{4.24}
u(x,t) = e^{i\omega t} \left(A\dn[\beta(x-ct),m]+ \frac{B\sqrt{1-m}}
{\dn[\beta(x-ct),m]} \right)\,,
\ee
is also an exact solution of the nonlocal Eq. (\ref{4.4}) provided
\be\label{4.25}
g > 0\,,~~ g A^2 =  \beta^2\,,~~B = \pm A\,,
~~c = [2-m \pm 6\sqrt{1-m}]\beta^2\,,
\ee
where the $\pm$ sign in $B = \pm A$ and in $v$ are correlated.

{\bf Solution XI: The Bion Solution}

The well known breather (also called bion) solution of the attractive local 
mKdV Equation 
\be\label{4.26}
u_t+u_{xxx}+6g u^2 u_x = 0\,,
\ee
is \cite{dj}
\be\label{4.27}
u(x,t) = -\frac{2}{\sqrt{g}}\frac{d}{dx} \tan^{-1} 
\bigg [\frac{c\sin(ax+bt+a_0)}{a\cosh(cx+dt+c_0)} \bigg ]\,,
\ee
provided
\be\label{4.28}
g > 0\,,~~b = a(a^2-3c^2)\,,~~d = c(3a^2-c^2)\,.
\ee
Here $a_0, c_0$ are arbitrary constants. It is then clear that 
the bion solution as given by Eq. (\ref{4.27}) is also the 
bion solution of the nonlocal mKdV Eq. (\ref{4.4}) provided $a_0 = c_0 = 0$.

{\bf Solution XII: Periodic Generalization of the Bion Solution}

It has been shown \cite{kks} that the periodic generalization of the bion 
solution of the local attractive mKdV Eq. (\ref{4.26}) is
\be\label{4.29}
u(x,t) = -\frac{2}{\sqrt{g}} \frac{d}{dx} \tan^{-1} 
\bigg [\alpha \sn(ax+bt+a_0, k)\dn(cx+dt+c_0, m) \bigg ]\,,
\ee
provided
\bea\label{4.30}
&&a^4 k = c^4(1-m)\,,~~\alpha = \frac{c}{a}\,, \nonumber \\
&&b = a[a^2(1+k)-3c^2(2-m)]\,,~~d = c[3a^2(1+k)-(2-m)c^2]\,. ~~~
\eea
As expected, in the limit $m \rightarrow 1, k \rightarrow 0$, the periodic
bion solution (\ref{4.29}) goes over to the bion solution (\ref{4.27}) and the 
relations between $c$ and $d$ as well as between $a$ and $b$ as given by 
Eq. (\ref{4.30}) go over to the one given in Eq. (\ref{4.28}). 

It is then clear that the periodic bion solution as given by Eq. (\ref{4.29}) 
is also the periodic bion solution of the nonlocal mKdV Eq. (\ref{4.4}) 
provided $a_0 = c_0 = 0$.

{\bf Solution XIII: Rational Solution}

It is easy to check that following the rational solution of the 
local mKdV Eq. (\ref{4.26}) \cite{dj}, the solution of the nonlocal 
mKdV Eq. (\ref{4.4}) is
\be\label{4.31}
\sqrt{g} u(x,t) = c - \frac{4c}{4c(x-6c^2 t)^2+1}\,.
\ee

{\bf Solution XIV}

It has been shown \cite{kksh} that there is another periodic solution of the
attractive ``local" mKdV Eq. (\ref{4.26}) and it is easy to check that it is 
also the solution of the nonlocal mKdV Eq. (\ref{4.4}) and is given by
\be\label{4.32}
u(x,t) = -\frac{2}{\sqrt{g}}\frac{d}{dx} \tan^{-1} 
\bigg [\alpha \sc(ax+bt, k) \dn(cx+dt, m) \bigg ]\,,
\ee
provided
\bea\label{4.33}
&&g > 0\,,~~(1-k) a^{4} = (1-m) c^4\,,~~\alpha^2 = -\frac{c}{a}\,, \nonumber \\
&&b = -a[a^2(2-k)+3c^2(2-m)]\,,~~d =- c[3a^2(2-k)+(2-m)c^2]\,. ~~~~
\eea

{\bf Solution XV}

Following \cite{kksh} it is easy to see that
\be\label{4.34}
u(x,t) = -\frac{2}{\sqrt{g}}\frac{d}{dx} \tan^{-1} 
\bigg [\alpha \sn(ax+bt, k)\sn(cx+dt, m) \bigg ]\,,
\ee
is an exact solution of the nonlocal mKdV Eq. (\ref{4.4}) provided  
\bea\label{4.35}
&&g > 0\,,~~k a^{4} = m c^4\,,~~\alpha^2 = \sqrt{km}\,, \nonumber \\
&&b = a[a^2(1+k)+3c^2(1+m)]\,,~~d = c[3a^2(1+k)+(1+m)c^2]\,. ~~~
\eea

{\bf Solution XVI}

Following \cite{kksh}, it is easy to see that another periodic solution of the
nonlocal mKdV Eq. (\ref{4.4}) is 
\be\label{4.37}
u(x,t) = -\frac{2}{\sqrt{|g|}} \frac{d}{dx} \tan^{-1} 
\bigg [\alpha \cn(ax+bt, k)\cn(cx+dt, m) \bigg ]\,,
\ee
provided
\bea\label{4.38}
&&g < 0\,,~~k(1-k) a^{4} = m(1-m) c^4\,,~~\alpha^2 = \frac{km}{(1-k)(1-m)}\,, 
\nonumber \\
&&b = a[a^2(1-2k)+3c^2(1-2m)]\,,~~d = c[3a^2(1-2k)+(1-2m)c^2]\,. ~~~~
\eea

Recently, we \cite{ks22c} have obtained four novel periodic and one hyperbolic 
solutions of local mKdV Eq. (\ref{4.26}) and shown that they can be 
re-expressed as superposed kink or pulse solutions. We now show that the 
nonlocal mKdV Eq. (\ref{4.4}) also admits these five 
superposed solutions.

{\bf Solution XVII}

Following \cite{ks22c} it is easy to see that
\be\label{4.39}
u(x,t) = \frac{A \dn(\xi,m) \cn(\xi,m)}{1+B\cn^2(\xi,m)}\,,~~B > 0\,,
\xi = \beta(x-ct)\,,
\ee
is an exact solution of the nonlocal mKdV Eq. (\ref{4.4}) provided
\bea\label{4.40}
&&g < 0\,,~~0 < m < 1\,,~~B = \frac{\sqrt{m}}{1-\sqrt{m}}\,,
\nonumber \\
&&c = -[1+m+6\sqrt{m}]\beta^2 < 0\,,
~~|g|A^2 = \frac{4 \sqrt{m} \beta^2}{(1-\sqrt{m})^2}\,.
\eea 
Note that this solution is not valid for $m = 1$, i.e. the nonlocal mKdV 
Eq. (\ref{4.4}) does not admit a corresponding hyperbolic solution.
Notice that for this solution $v < 0$.

On using the identity (\ref{7}) one can then rewrite the periodic 
pulse solution (\ref{4.39}) as a superposition of a periodic kink and
a periodic antikink solution, i.e. 
\be\label{4.41}
u(x,t) = \sqrt{\frac{2m}{|g|}} \beta 
[\sn(\xi +\Delta, m) -\sn(\xi- \Delta, m)]\,,~~\xi = \beta(x-ct)\,.
\ee
Here $\Delta$ is defined by $\sn(\sqrt{m}\Delta,1/m) = \pm m^{1/4}$,
where use has been made of the identity (\ref{3.20}). 

{\bf Superposed Solution XVIII}

Following \cite{ks22c} it is easy to show that the nonlocal mKdV 
Eq. (\ref{4.4}) admits the periodic kink solution
\be\label{4.42}
u(x,t) = \frac{A \sn(\xi,m)}{1+B\cn^2(\xi,m)}\,,~~B > 0\,,
\ee
provided
\bea\label{4.43}
&&g < 0\,,~~0 < m < 1\,,~~B = \frac{\sqrt{m}}{1-\sqrt{m}}\,,
\nonumber \\
&&c = [6\sqrt{m}-(1+m)]\beta^2\,,~~ |g|A^2 = 4\sqrt{m} \beta^2\,.
\eea 
Note that this solution does not exist for $m = 1$, i.e. the corresponding
hyperbolic solution does not exist. 

On using the identity (\ref{6}),  the periodic solution (\ref{4.42}) 
can be re expressed as
\be\label{4.44}
u(x,t) = i \sqrt{\frac{m}{|g|}} \beta  
[\sn(\xi +\Delta, m)+\sn(\xi -\Delta, m)]\,,~~\xi = \beta(x-ct)\,.
\ee
Here $\Delta$ is defined by $\sn(\sqrt{m}\Delta,1/m) = \pm m^{1/4}$,
where use has been made of the identity (\ref{3.20}).

It is worth noting that for both the superposed solutions XVIII and XIX 
of the nonlocal repulsive mKdV Eq. (\ref{4.4}), the value of $B$ as well as
$g$ are the same.

{\bf Solution XIX}

Following \cite{ks22c} it is easy to show that another 
periodic solution 
to the nonlocal mKdV Eq. (\ref{4.4}) is
\be\label{4.45}
u(x,t) = \frac{A \sn(\xi,m) \cn(\xi,m)}{1+B\cn^2(\xi,m)}\,,
\ee
provided
\bea\label{4.46}
&&0 < m < 1\,,~~B = \frac{1-\sqrt{1-m}}{\sqrt{1-m}}\,,~~
c = (2-m-6\sqrt{1-m})\beta^2\,, \nonumber \\
&&g < 0\,,~~|g|A^2 = \frac{4(1-\sqrt{1-m})^2 \beta^2}{\sqrt{1-m}}\,.
\eea

On using the identity (\ref{11}), the periodic solution (\ref{4.45})  
can be re-expressed as a 
superposition of the two periodic pulse solutions, i.e.
\be\label{4.47}
u(x,t) = \frac{\beta}{\sqrt{|g|}} \big (\dn[\xi -K(m)/2,m] 
- \dn[\xi +K(m)/2,m] \big )\,, ~~\xi = \beta(x-ct)\,.
\ee

{\bf Solution XX}

Following \cite{ks22c}, yet another periodic solution to the nonlocal mKdV 
Eq. (\ref{4.4}) is
\be\label{4.48}
u(x,t) = \frac{A \dn(\xi,m)}{1+B\cn^2(\xi,m)}\,,
\ee
provided
\bea\label{4.49}
&&0 < m < 1\,,~~ B = \frac{1-\sqrt{1-m}}{\sqrt{1-m}}\,,~~g > 0\,,
\nonumber \\
&&c = [2-m+6\sqrt{1-m}]\beta^2\,,~~g A^2 = \frac{4}{\sqrt{1-m}} \beta^2\,.
\eea
Thus for this solution $c > 0$.

On using the identity (\ref{10}) 
the periodic solution (\ref{4.48}) can be
re-expressed as superposition of two periodic pulse solutions, i.e.
\be\label{4.50}
u(x,t) = \beta  
[\dn(\beta x +K(m)/2, m)+\dn(\beta x -K(m)/2, m)]\,.
\ee

{\bf Superposed Solution XXI}

Following \cite{tan, ks22c} it is easy to check that the 
nonlocal mKdV Eq. (\ref{4.4}) admits a hyperbolic pulse solution
\be\label{4.51}
\psi(x,t) = 1-\frac{A}{B+\cosh^2(\xi)}\,,~~B > 0\,,~~\xi = \beta(x-ct)\,,
\ee
provided 
\be\label{4.52}
g < 0\,,~~\sqrt{|g|} A = 2\sqrt{B(B+1)}\beta\,,~~
\beta^2 = \frac{4(B+1)}{(B+2)^2} < 1\,,~~c = 4\beta^2 -6\,.
\ee
On comparing with the hyperbolic identity (\ref{14}),
the solution (\ref{4.51}) can be
re-expressed as the superposition of a (hyperbolic) kink and an 
antikink solution, i.e.
\be\label{4.53}
u(x,t) = 1-\frac{\beta}{\sqrt{|g|}}[\tanh(\xi+\Delta) -\tanh(\xi-\Delta)]\,,
\ee
where $\xi = \beta(x-ct)$ while $\sinh(\Delta) = \sqrt{B}$.

\end{document}